\documentclass[aps,prl,reprint,twocolumn,amsmath,amssymb,showpacs,superscriptaddress, floatfix, longbibliography]{revtex4-2}

\usepackage[most]{tcolorbox}
\usepackage{amsmath, amssymb}
\usepackage{graphicx}
\usepackage[margin=1in]{geometry}
\usepackage{array}
\usepackage{color}
\usepackage{textcomp} 
\usepackage{mathrsfs,amsmath}
\usepackage{mathtools}
\usepackage{graphicx}
\graphicspath{{Figures/}} 
\usepackage{dcolumn}
\usepackage{lipsum}
\usepackage[mathlines]{lineno}
\usepackage{hyperref}
\usepackage{xcolor}
\hypersetup{
    colorlinks,
    linkcolor={red!50!black},
    citecolor={blue!50!black},
    urlcolor={blue!80!black}
}
\usepackage{bm}
\usepackage{float}
\usepackage{epstopdf}
\usepackage{soul}
\usepackage{epsfig}
\usepackage{bbold}
\usepackage{braket}
\usepackage{physics}
\usepackage{gensymb}
\usepackage{amsmath,amssymb}
\usepackage{caption}
\usepackage{ragged2e}

\begin{document}

\title{Vortex Light at the Nanoscale: Twists, Spins, and Surprises – A Review}
\author{Kayn A. Forbes}
\email{K.Forbes@uea.ac.uk}

\affiliation{School of Chemistry, University of East Anglia, Norwich Research Park, Norwich NR4 7TJ, United Kingdom}

\begin{abstract}
For over three decades, the study of optical vortex beams carrying orbital angular momentum (OAM) has been at the forefront of optics, driven by fundamental questions about optical momentum as well as diverse applications in quantum information, communications, and optical manipulation. Most work has focused on paraxial beams, whose transverse fields are accurately described by conventional wave optics and the Stokes formalism. By contrast, when light is confined to the nanoscale and tightly focused beyond the paraxial regime, vortex beams exhibit complex electromagnetic structures that transcend these conventional models. In this deeply non-paraxial regime, the resulting fields display rich and often counterintuitive behaviour, opening new perspectives on light–matter interactions. This review unifies the emerging physics of nanoscale optical vortices by developing a coherent theoretical framework and offering a critical synthesis of recent advances, guiding readers toward a deeper understanding and stimulating future work in this rapidly evolving field.
\end{abstract}

\maketitle
\onecolumngrid
\section{Introduction} \label{I}

The behavior and properties of light have long been recognised as far richer than those described by the ubiquitous plane-wave approximation. However, the 1992 work by Les Allen and colleagues on the orbital angular momentum (OAM) of Laguerre-Gaussian (LG) modes – now broadly termed `optical vortices' or `twisted light' – catalysed a transformative shift in optics research \cite{allen1992orbital, franke202230}. In the three decades since, the field of structured light has flourished, opening up new frontiers in both classical and quantum optics, and enabling a wealth of technological applications \cite{andrews2011structured, shen2019optical, forbes2019structured, angelsky2020structured, yang2021optical, forbes2021structured, he2022towards}. Parallel to this, nano-optics has emerged as a mature and powerful area of modern optics, concerned with the behavior of light confined to subwavelength scales, often mediated by evanescent waves and near-field interactions \cite{novotny2012principles, benisty2022introduction}. In both structured and nanoscale optics, light exhibits behaviour that deviates dramatically from the textbook norms.

This review brings these two threads together, focusing on the \emph{nano-optics of structured beams} – particularly optical vortices – and how tight focusing drives entirely new regimes of behaviour. When these intricate beams are compressed to nanoscale dimensions, their hidden structure unfolds, giving rise to phenomena such as transverse spin and chiral unpolarized light that push the boundaries of our fundamental understanding of optical fields. While comprehensive texts and reviews exist on structured light \cite{andrews2011structured, shen2019optical, forbes2019structured, angelsky2020structured, yang2021optical, forbes2021structured, he2022towards} and on nano-optics \cite{novotny2012principles, benisty2022introduction} individually, this review instead focuses on their intersection, aiming to provide a robust theoretical foundation that gives the reader clear physical insight into the most recent cutting-edge developments emerging from their combination. We highlight the unique features that arise specifically from non-paraxiality – where the tight focusing of structured beams gives rise to spin–orbit interactions, longitudinal field components, and rich polarisation structures that are both quantitatively and qualitatively distinct from those in the paraxial regime, rather than being generic consequences of high numerical aperture (NA) optics. We do not cover the optical physics of tight focusing or nanoscale vortex beam generation, which lie beyond the scope of this review but remain active research areas \cite{chen2012tight, garoli2016optical, peatross2017vector, kozawa2021small, prinz2023orbital, chen2024highly}

We begin the review in Section~\ref{II} by providing a brief background to structured light and nano-optics followed by an in-depth explanation in Section~\ref{III} of the theory we will use throughout the Review to describe a focused vortex beam. We then systematically work our way through the energy (Section~\ref{IV}); momentum (Section~\ref{V}); angular momentum (AM) (Section~\ref{VI}); and chirality (Section~\ref{VII}) properties of focused vortex beams. We conclude the review with discussions of vortex beam atom optics in Section~\ref{VIII} and the future of the field in Section~\ref{IX}. 

\section{Structured Light and Nano-optics} \label{II}

Unstructured light refers to beams with spatially homogeneous degrees of freedom (polarisation, amplitude, phase), with the paraxial fundamental Gaussian mode being the archetypal example. In contrast, structured light allows for the tailoring of these degrees of freedom to create spatially and temporally varying (inhomogeneous) distributions \cite{forbes2021structured}. As introduced earlier, the ability to design light modes with desired electromagnetic field structures has made structured light a key tool in numerous important applications. Most of these applications operate in the paraxial regime of optics, where the physics is well described by the notion of light as a transverse electromagnetic wave (with respect to the direction of propagation), and fully captured by the well-known Stokes vector description. The optical properties of paraxial beams of light are well-known and take on relatively simple forms: Box 1.

What distinguishes nano-optics from classical optics is its focus on spatially confined electromagnetic fields \cite{novotny2012principles}. This confinement renders the light in nano-optics as non-paraxial, in stark contrast to the paraxial fields of classical optics. Non-paraxial beams or fields are not purely transverse to the direction of propagation, acquiring longitudinal components along the propagation axis – see Figure~\ref{fig:1}. The relative strength of these longitudinal components increases as the field becomes more tightly confined. For instance, a tightly focused beam exhibits a much larger longitudinal component compared to a weakly focused one. In contrast to paraxial beams, which can be described using the familiar two-dimensional (2D) Stokes vector – because the fields are polarised strictly in the plane orthogonal to propagation – non-paraxial light requires a three-dimensional (3D) Stokes vector for its description, due to the presence of components in all directions \cite{alonso2023geometric}.


It is instructive to use the language of Fourier optics – or, equivalently,  Heisenberg's uncertainty principle – to understand why electromagnetic fields  are inherently three-dimensionally polarised. Consider a beam propagating along  the \( z \)-axis and initially described as a purely transverse wave with  \(|\mathbf{k}| = k_z\). Confining the beam in real space broadens its spectrum  in \(\mathbf{k}\)-space, so that \( k_x \) and \( k_y \) components inevitably  appear (classically a Fourier result and in quantum terms  \(\Delta k_i \geq 1/(2\Delta r_i)\)). Due to the transversality condition \(\mathbf{E}\cdot\mathbf{k} = 0\), these off-axis components force the electric field to acquire additional components, including a longitudinal term \( E_z \). This directly explains why spatial confinement is essential for the emergence of non-paraxial field components. Moreover, even an \( x \)-polarised paraxial beam will, under sufficient confinement, develop not only a longitudinal field component \( E_z \), but also a transverse component \( E_y \) orthogonal to the initial polarisation to ensure transversality (see Figure~\ref{fig:1}b). In the next Section, we identify these as the first-order longitudinal and second-order transverse field components, respectively.

A consequence of the paraxial approximation is that the optical properties of free-space electromagnetic fields are dual-symmetric: the electric and magnetic field contributions are identical. For example, the energy density of a paraxial beam is given by: 
\begin{align}
    u_\text{Paraxial}=\frac{\epsilon_0}{4}\bigl(|\mathbf{E}|^2+c^2|\mathbf{B|}^2\bigr)=\frac{\epsilon_0}{2}|\mathbf{E}|^2=\frac{\epsilon_0c^2}{2}|\mathbf{B}|^2.
    \label{eq:1}
\end{align}
However, for non-paraxial electromagnetic fields, the dual-asymmetric contributions of the electric and magnetic fields to the energy density yield:
\begin{align}
    u=\frac{\epsilon_0}{4}\bigl(|\mathbf{E}|^2+c^2|\mathbf{B|}^2\bigr)=u_\text{E} + u_\text{B},
    \label{eq:2}
\end{align}
where $u_\text{E} = \frac{\epsilon_0}{2}|\mathbf{E}|^2 \neq u_\text{B} = \frac{\epsilon_0 c^2}{2}|\mathbf{B}|^2$. This is why the 2D Stokes description can be based purely on the electric field alone: the electric and magnetic contributions are equivalent. In stark contrast, the non-paraxial fields of nano-optics possess dual-asymmetric optical properties \cite{bliokh2013dual}: the individual electric and magnetic fields contributions are, in general, distinctly different. This critical distinction between paraxial and non-paraxial fields is worth bearing in mind as we progress through the review, as it has profound implications for light-matter interactions.

\begin{tcolorbox}[
    colback=gray!2,                  
    colframe=gray!50!black,                 
    coltitle=white,                 
    boxrule=1pt,                    
    title={\normalsize\bfseries Box 1 – Optical properties of paraxial vortex beams},
    fonttitle=\bfseries,            
    arc=2pt,                        
    enhanced,
    sharp corners=south,            
    width=\textwidth,
    drop shadow southeast,          
    top=6pt, bottom=6pt,            
    left=6pt, right=6pt,            
    boxsep=4pt                      
]

\renewcommand{\arraystretch}{1.3} 
\setlength{\tabcolsep}{0.5em}     

\begin{tabular}{|p{0.48\textwidth}|p{0.48\textwidth}|}

\hline

\textbf{Energy Density} \newline
The time-averaged energy density of a paraxial Laguerre–Gaussian (LG) beam is proportional to its intensity $I_\text{LG}$, with the field profile given by the mode $\psi^{\ell,p}_\text{LG}$ in Eq.~\eqref{eq:4}:
\[
u = \frac{\epsilon_0}{2}|\mathbf{E}|^2 = \frac{\epsilon_0}{2}\left|\psi^{\ell,p}_{\text{LG}}\right|^2 = \frac{I_\text{LG}}{c}
\]
\[
\textit{per photon:} \quad \hbar\omega
\]
\includegraphics[width=\linewidth]{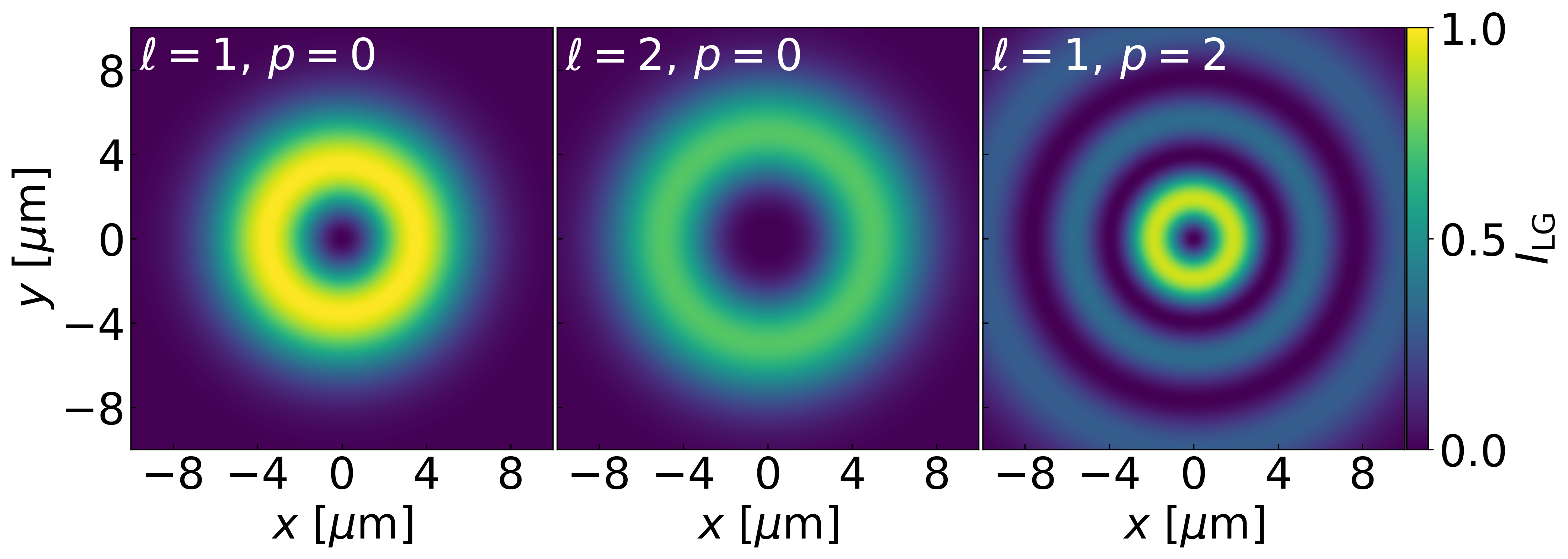} &

\textbf{Helicity density} \newline
For monochromatic beams, the helicity density $h$ is proportional to the optical chirality $C$ and reflects the degree of ellipticity in the paraxial polarisation. It reaches its maximum magnitude for circular polarisation, where $\sigma = \pm1$ denotes left- or right-handedness:
\[
h=-\frac{\epsilon_0}{2\omega}\text{Im}(\mathbf{E}^*\cdot\mathbf{B}) = \sigma \frac{I_\text{LG}}{c\omega}
\]
\[
\textit{per photon:} \quad \sigma\hbar
\]
\includegraphics[width=\linewidth]{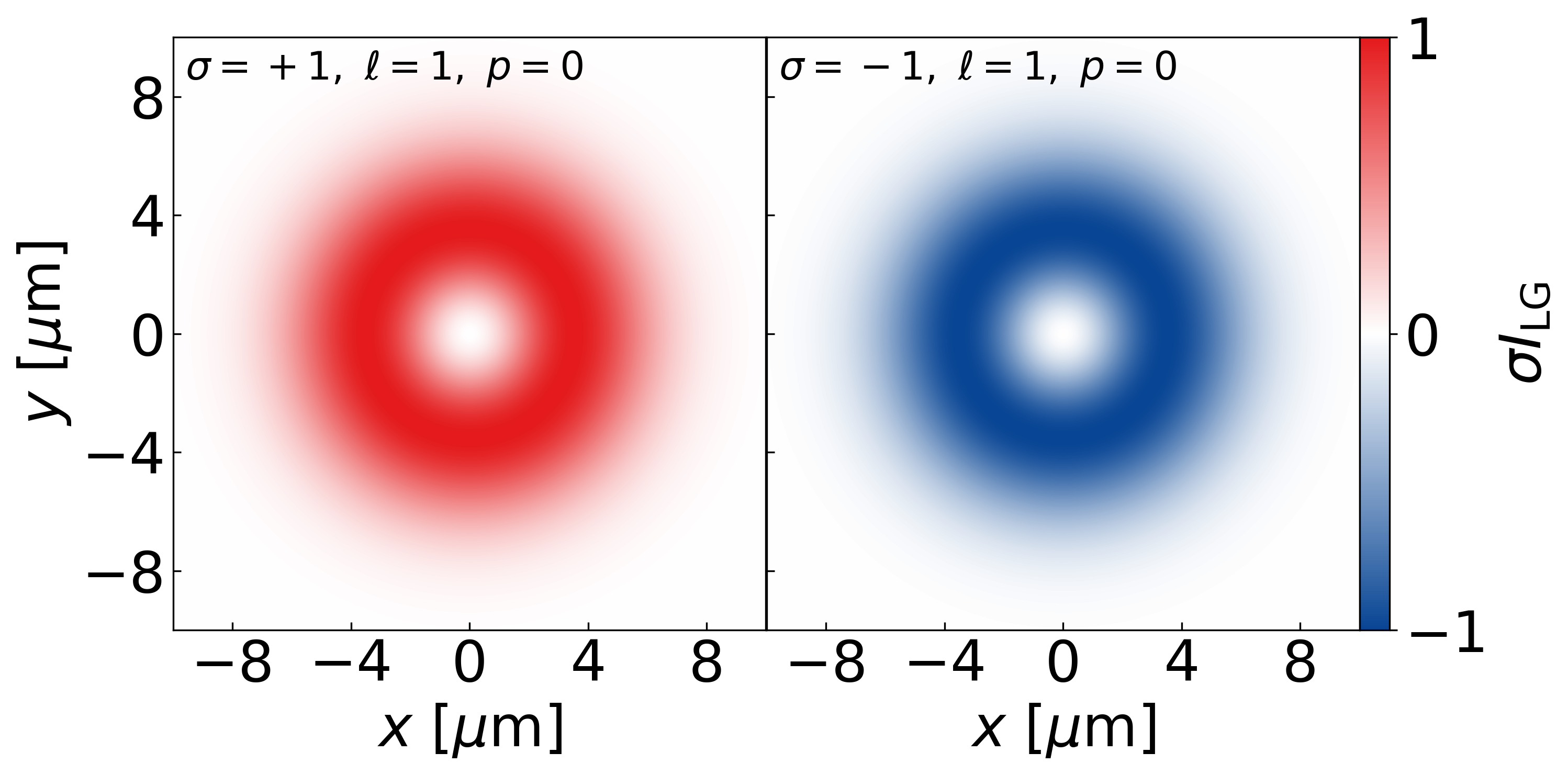} \\
\hline

\textbf{Canonical (orbital) momentum density} \newline
The canonical (or orbital) momentum density for a $z$-propagating LG beam is given by:
\[
\mathbf{p}_\text{o} = \frac{\epsilon_0}{2\omega}\text{Im}(\mathbf{E}^*\cdot\nabla\mathbf{E})=\frac{I_\text{LG}}{c\omega}\Bigl(k\mathbf{\hat{z}}+\frac{\ell}{r}\mathbf{\hat{\phi}}\Bigr)
\]
\[
\textit{per photon:} \quad \hbar k \mathbf{\hat{z}}
\]
\includegraphics[width=\linewidth]{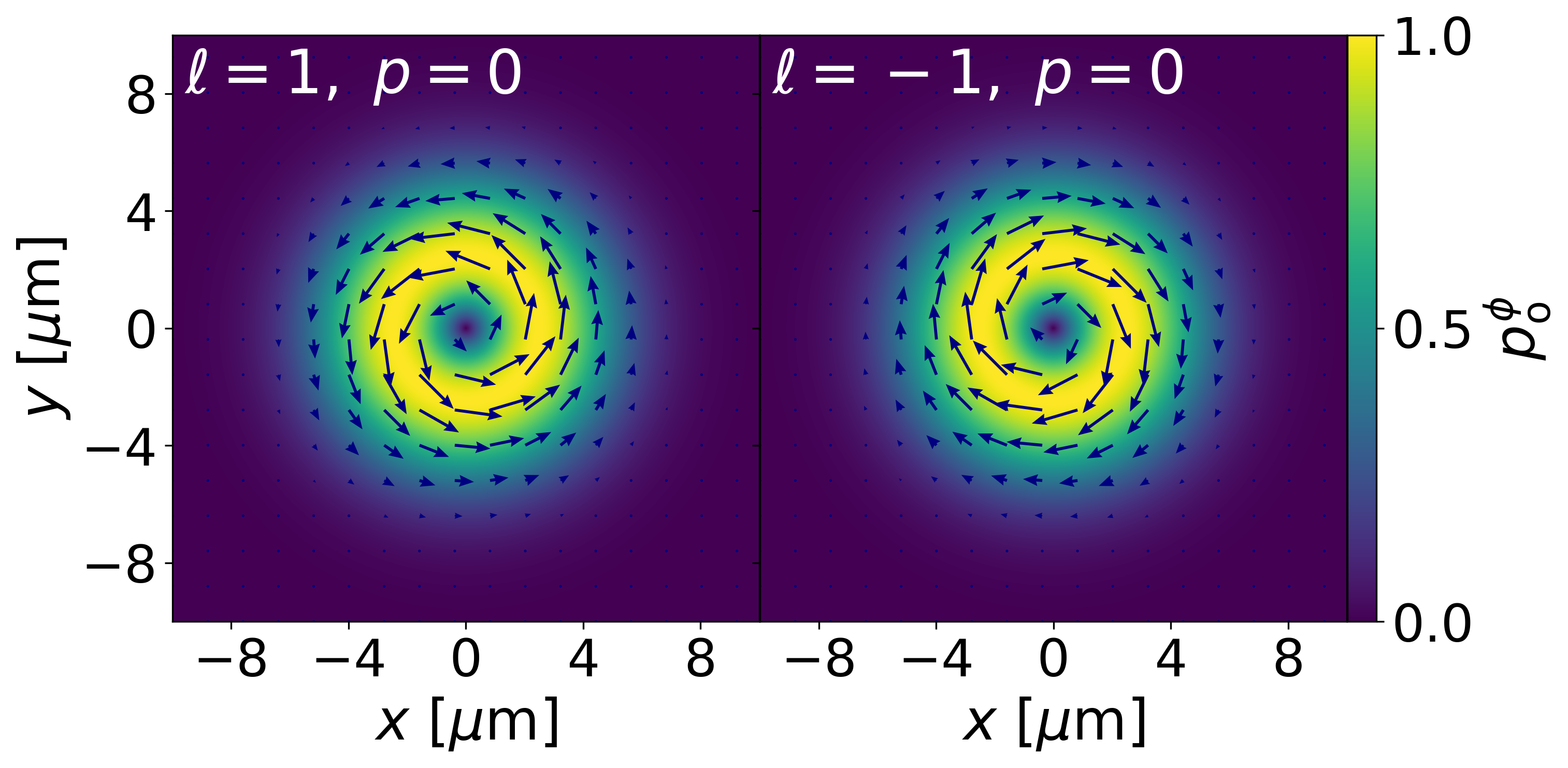} &

\textbf{Spin momentum density} \newline
Belinfante’s spin momentum density, the curl of the spin angular momentum (SAM), reflects the local polarisation structure and is purely azimuthal for a circularly polarised (CP) paraxial LG beam:
\[
\mathbf{p}_\text{s} = \frac{\epsilon_0}{4\omega}\nabla\times\text{Im}(\mathbf{E}^*\times\mathbf{E})=-\frac{\partial}{\partial r}\frac{\sigma I_\text{LG}}{2c\omega}\mathbf{\hat{\phi}}
\]
\[
\textit{per photon:} \quad 0
\]
\includegraphics[width=\linewidth]{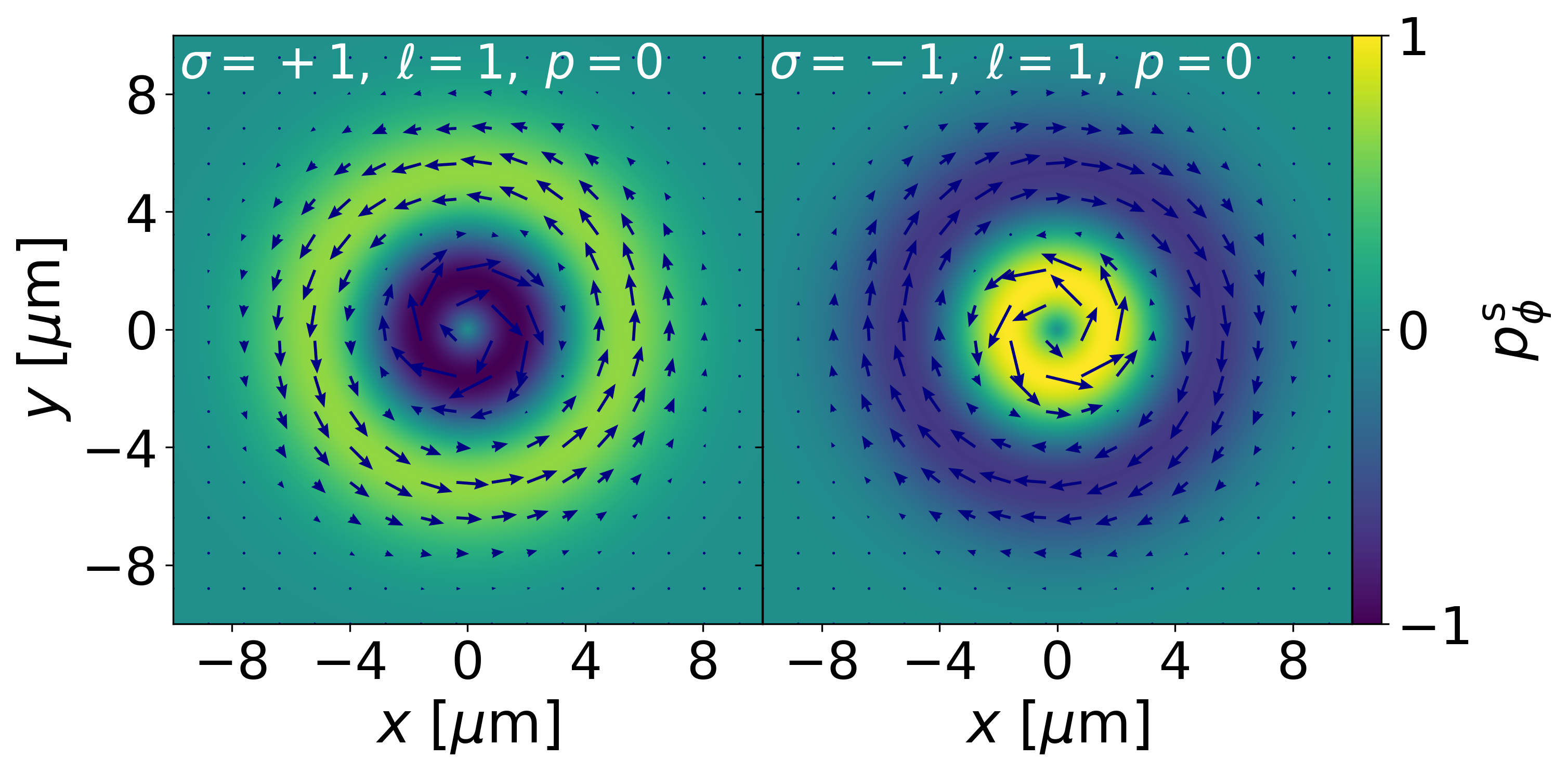} \\
\hline

\textbf{Spin angular momentum density} \newline
The spin angular momentum density of a paraxial beam is equivalent to the helicity density directed along the propagation axis:
\[
\mathbf{s} = \frac{\epsilon_0}{2\omega}\text{Im}(\mathbf{E}^*\times\mathbf{E})=\sigma\frac{I_\text{LG}}{c\omega}\mathbf{\hat{z}}
\]
\[
\textit{per photon:} \quad \sigma \hbar \mathbf{\hat{z}}
\]
&

\textbf{Orbital angular momentum density} \newline
The $\ell$-dependent orbital angular momentum density is directed along the propagation axis and its spatial distribution matches the beam intensity profile:
\[
\mathbf{l} = \mathbf{r}\times\mathbf{p}_{\text{o}}=\frac{I_\text{LG}}{c\omega}\Big(-kr\mathbf{\hat{\phi}}+\ell\mathbf{\hat{z}}\Bigr)
\]
\[
\textit{per photon:} \quad \ell \hbar \mathbf{\hat{z}}
\]
\vspace{-1em} 
\\

\hline
\end{tabular}

\end{tcolorbox}

\section{Analytical description of optical vortex beams at the nanoscale} \label{III} 

Optical vortices is a generic term that encompasses several modes carrying OAM, such as LG, Bessel, Bessel-Gauss, and Mathieu modes \cite{andrews2012angular, shen2019optical}. Such OAM-carrying modes have found diverse applications in fields such as optical trapping and manipulation, quantum information processing, microscopy, material processing, and studies of fundamental light-matter interactions \cite{franke202230, shen2019optical, forbes2021orbital, franke202230, porfirev2023light}. With respect to many applications, the important factor in all such modes is the azimuthal phase $\text{e}^{i\ell\phi}$, where $\ell\in\mathbb{Z}$ is the topological charge. Modes with $\text{e}^{i\ell\phi}$ are eigenfunctions of the OAM operator $l_z=-i\hbar\frac{\partial}{\partial\phi}$ giving the eigenvalue $\ell\hbar$ of OAM per photon.  In this review we will use the LG modes to provide a theoretical narrative due the fact they are the most prevalent vortex modes utilised in experiments and applications. 

A paraxial beam is defined as one whose electromagnetic field is fully transverse to the propagation direction. This condition is typically realised for beams with a waist $w_0$ much larger than the wavelength $w_0\gg\lambda$. The electric field of a paraxial monochromatic LG beam propagating along $z$ is 
\begin{align}
    \mathbf{E}^\text{T0}_{\text{LG}} = \bigl( \alpha \mathbf{\hat{x}} + \beta \mathbf{\hat{y}} \bigr) \psi_{\text{LG}}^{\ell,p}\,,
    \label{eq:3}
\end{align}
where $\psi_{\text{LG}}^{\ell,p}$ is a normalised solution to the paraxial wave equation in cylindrical coordinates~\cite{andrews2012angular}
\begin{align}
\psi_{\text{LG}}^{\ell,p} &=  \nonumber \sqrt{\frac{2p!}{{\pi w_{0}^2}(p+|\ell|)!}}\frac{w_0}{w[z]}
\Biggr(\frac{\sqrt{2}r}{w[z]}\Biggr)^{|\ell|}
L_{p}^{|\ell|}\Biggr[\frac{2r^2}{w^2[z]}\Biggr] \text{exp}(-r^2/w^2[z])
\\ &    \times \text{exp}i(kz+\ell\phi+kr^2/2R[z] -\omega t
- (2p + |\ell| + 1)\zeta [z]),.
    \label{eq:4}
\end{align}
In the above $\alpha$ and $\beta$ are the (generally complex) Jones vector coefficients: $\abs{\alpha}^2+\abs{\beta}^2=1$, $\ell \in \mathbb{Z}$ and $p \in \mathbb{Z^+}$  are the topological charge and radial index of the LG beam, respectively, $R[z]$ is the wavefront curvature, $\zeta[z]$ is the Gouy phase, $L_{p}^{|\ell|}\bigr[\frac{2r^2}{w^2[z]}\bigr]$ is the generalized Laguerre polynomial, square brackets are reserved for functional dependencies, all other symbols have their usual meaning, and the notation `T0' in Eq.~\eqref{eq:3} will be explained below. 

Although perhaps not widely recognised, it has long been established that paraxial beams do not satisfy Maxwell's equations~\cite{lax1975maxwell}. For instance, Eq.~\eqref{eq:3} is not divergence-free, as required by Gauss's law: $\nabla \cdot \mathbf{E}^\text{T0}_{\text{LG}} \neq 0$. In fact, it has been demonstrated~\cite{lekner2002polarization} that for any real electromagnetic beam, 
\(\nabla_\perp \cdot \mathbf{E} = 0\) and \(\nabla_\perp \cdot \mathbf{B} = 0\) cannot be satisfied simultaneously. 
More generally, both transverse divergences are nonzero, i.e., 
\[
\nabla_\perp \cdot \mathbf{E} \neq 0 \quad \text{and} \quad \nabla_\perp \cdot \mathbf{B} \neq 0.
\]
Nonetheless, under paraxial conditions (weak-focusing or collimated), Eq.~\eqref{eq:3} provides a reasonable approximation of the electric field. Beams that are strongly focused, however, become non-paraxial, acquiring significant longitudinal and higher-order transverse electromagnetic field components. In such cases, paraxial descriptions of the electromagnetic fields (Eq.~\eqref{eq:3}) are clearly inadequate, and alternative methods – either analytical or numerical – should be used~\cite{peatross2017vector}. Thanks to the strict boundary conditions required by numerical techniques based on diffraction integrals, these methods can offer closed-form solutions with high quantitative accuracy. Moreover, numerical techniques also allow for incorporation of experimental parameters, such as the numerical aperture, focal length, and filling factor. Analytical approaches, in contrast, provide greater physical insight into the behavior of focused beams, revealing how the higher-order electromagnetic field components contribute to optical properties such as energy, momentum, and angular momenta.

To provide a transparent explanation of the electromagnetic origin of the behaviors of structured light at the nanoscale in this review, we adopt an analytical approach. Readers interested in numerical and analytical methods can refer to~\cite{peatross2017vector} for a broader overview. Our method is based on the fact that, under focusing, higher-order vector corrections to paraxial electromagnetic fields must be considered, as they are generated in proportion to the paraxial parameter. These higher-order field components are obtained by solving Maxwell's equations iteratively. For Gaussian-type beams, the paraxial parameter is given by $1/kw$. As a result, the paraxial field, such as the one described in Eq.~\eqref{eq:3}, is referred to as the zeroth-order transverse field T0 with respect to $1/kw$.

The first non-paraxial correction is the first-order longitudinal field `L1', followed by the second-order transverse field `T2', and so on. Longitudinal (transverse) components are always odd(even)-order in the paraxial parameter. The analytical method of describing a non-paraxial beam we will use involves taking the zeroth-order description of the electromagnetic field of Eq.~\eqref{eq:3}, which has a nonzero divergence, and using Maxwell's equations in an iterative process to generate the higher-order terms. 
The unknown term `L1' is obtained imposing the Gauss's law on $ \mathbf{E}^\text{T0} + \mathbf{E}^\text{L1}$:
\begin{align}
    \nabla \cdot \mathbf{E}^\text{T0 + L1} = \nabla_{\perp} \cdot \mathbf{E}^\text{T0} + \frac{\partial }{\partial z}E^\text{L1}_z = 0\,,
    \label{eq:5}
\end{align}
which rearranging leads to,
\begin{align}
    E^\text{L1}_z =-\int \nabla_{\perp} \cdot \mathbf{E}^\text{T0} {\partial z}.
    \label{eq:6}
\end{align}
In order to secure an analytical result for $E_z$, it is at this point the approximation is made that the variation in $z$ is dominated by the $\text{e}^{ikz}$ phase factor, which leads to~\cite{adams2018optics}
\begin{align}
    E^\text{L1}_z \approx \frac{i}{k} \nabla_{\perp} \cdot \mathbf{E}^\text{T0}\,.
    \label{eq:7}
\end{align}
This shows that the first-order longitudinal electric field component $E^\text{L1}_z$ is directly proportional to the transverse gradient of the zeroth-order transverse field. Applying Gauss's law to the obtained field, obviously results in a zero divergence ($\nabla \cdot (\mathbf{E}^\text{T0}_{\text{LG}} + \mathbf{\hat{z}}{E}^\text{L1}_{\text{LG}}) = 0$), as required by the Maxwell's equations. 
Continuing the iteration, the field $\mathbf{E}^\text{T0}_{\text{LG}} + \mathbf{E}^\text{L1}_{\text{LG}}$ is inserted into the Faraday's Law to generate the magnetic field, up to the second order in $1/kw$ ($\mathbf{B}^\text{T0}_{\text{LG}} + \mathbf{B}^\text{L1}_{\text{LG}} + \mathbf{B}^\text{T2}_{\text{LG}}$). Finally,  $\mathbf{B}^\text{T0 + L1}_{\text{LG}}$ is plugged in the Maxwell-Ampere law, yielding the second-order transverse electric field  $\mathbf{E}^\text{T0}_{\text{LG}} + \mathbf{E}^\text{L1}_{\text{LG}} + \mathbf{E}^\text{T2}_{\text{LG}}$. With a single application of the iteration, both the electric and magnetic fields are obtained to second order in the parameter $1/kw$, resulting in the explicit expression for the electric field:

\begin{align}
    \mathbf{E}^\text{T0+L1+T2}_{\text{LG}} &=  \psi_{\text{LG}}^{\ell,p}\, \Bigl\{\overbrace{\alpha \mathbf{\hat{x}} + \beta \mathbf{y}}^{\text{T0}} + \overbrace{\mathbf{\hat{z}}\frac{i}{k} \Bigl[\alpha \bigl( \gamma \cos\phi - \frac{i\ell}{r} \sin\phi \bigr) + \beta \bigl( \gamma \sin\phi + \frac{i\ell}{r} \cos\phi \bigr) \Bigr]}^{\text{L1}} \nonumber \\
    &+ \frac{1}{k^2} \mathbf{\hat{x}} \Bigl[2\alpha\sin\phi\cos\phi\Bigl(\frac{i\ell}{r^2}-\frac{i\ell\gamma}{r}\Bigr) + \alpha\cos^2\phi\Bigl(\frac{\ell^2}{r^2}-\frac{\gamma}{r}\Bigr)-\alpha\sin^2\phi\bigl\{\gamma^{'} + \gamma^2\bigr\} \nonumber \\ & + \beta\sin\phi\cos\phi\Bigl(\gamma^{'} + \gamma^2 - \frac{\gamma}{r} + \frac{\ell^2}{r^2} \Bigr) + \beta \cos^2\phi \Bigl(\frac{i\ell\gamma}{r} - \frac{i\ell}{r^2} \Bigr) + \beta \sin^2\phi \Bigl(\frac{i\ell}{r^2} - \frac{i\ell\gamma}{r} \Bigr) \Bigr) \nonumber \\
    &+ \frac{1}{k^2} \mathbf{\hat{y}} \Bigl[\alpha\sin\phi\cos\phi\Bigl(\gamma^{'}+\gamma^2-\frac{\gamma}{r}+\frac{\ell^2}{r^2}\Bigr) + \alpha\cos^2\phi\Bigl(\frac{i\ell\gamma}{r} - \frac{i\ell}{r^2}\Bigr) + \alpha\sin^2\phi\Bigl(\frac{i\ell}{r^2}-\frac{i\ell\gamma}{r}\Bigr) \nonumber \\
    &+ 2\beta\sin\phi\cos\phi\Bigl(\frac{i\ell\gamma}{r} - \frac{i\ell}{r^2}\Bigr) -\beta\cos^2\phi\Bigl(\gamma^{'} + \gamma^{2}\Bigr) + \beta\sin^2\phi\Big(\frac{\ell^2}{r^2}-\frac{\gamma}{r}\Bigr)\Bigr]\Bigr) \,,
    \label{eq:8}
\end{align}

where 
\begin{align}
    \gamma\qty(\vb{r}) &= \frac{|\ell|}{r}-\frac{2r}{w^2}+ \frac{ikr}{R[z]} -\frac{4r}{w^2}\frac{L^{|\ell|+1}_{p-1}}{L^{|\ell|}_{p}}\,,
    \label{eq:9}
\end{align}
which comes from the derivative of the field over the radial coordinate $r$: $\frac{\partial}{\partial r} \psi_{\text{LG}}^{\ell,p} = \gamma \psi_{\text{LG}}^{\ell,p}$. Note that $\gamma^{'} = \frac{\partial}{\partial r} \gamma$ and for $p=0$, ${L^{|\ell|+1}_{p-1}}=0$.

The longitudinal and the higher-order transverse components of the field in Eq.~\eqref{eq:8} are of order one and two in the paraxial parameter, respectively, so that in a paraxial beam they are negligible compared to the zeroth order term T0. In tightly focused beams, as $w$ becomes comparable in size with $\lambda$, the paraxial parameter $1/kw\longrightarrow1$, meaning that the paraxial approximation is broken and the higher-order fields thus become appreciable with respect to the T0 term. By fixing the beam waist to wavelength ratio, Eq.~\eqref{eq:8} can model both a paraxial and non-paraxial LG mode of any order $(\ell,p)$.  

\begin{figure}[!ht]
\includegraphics[width = \linewidth]{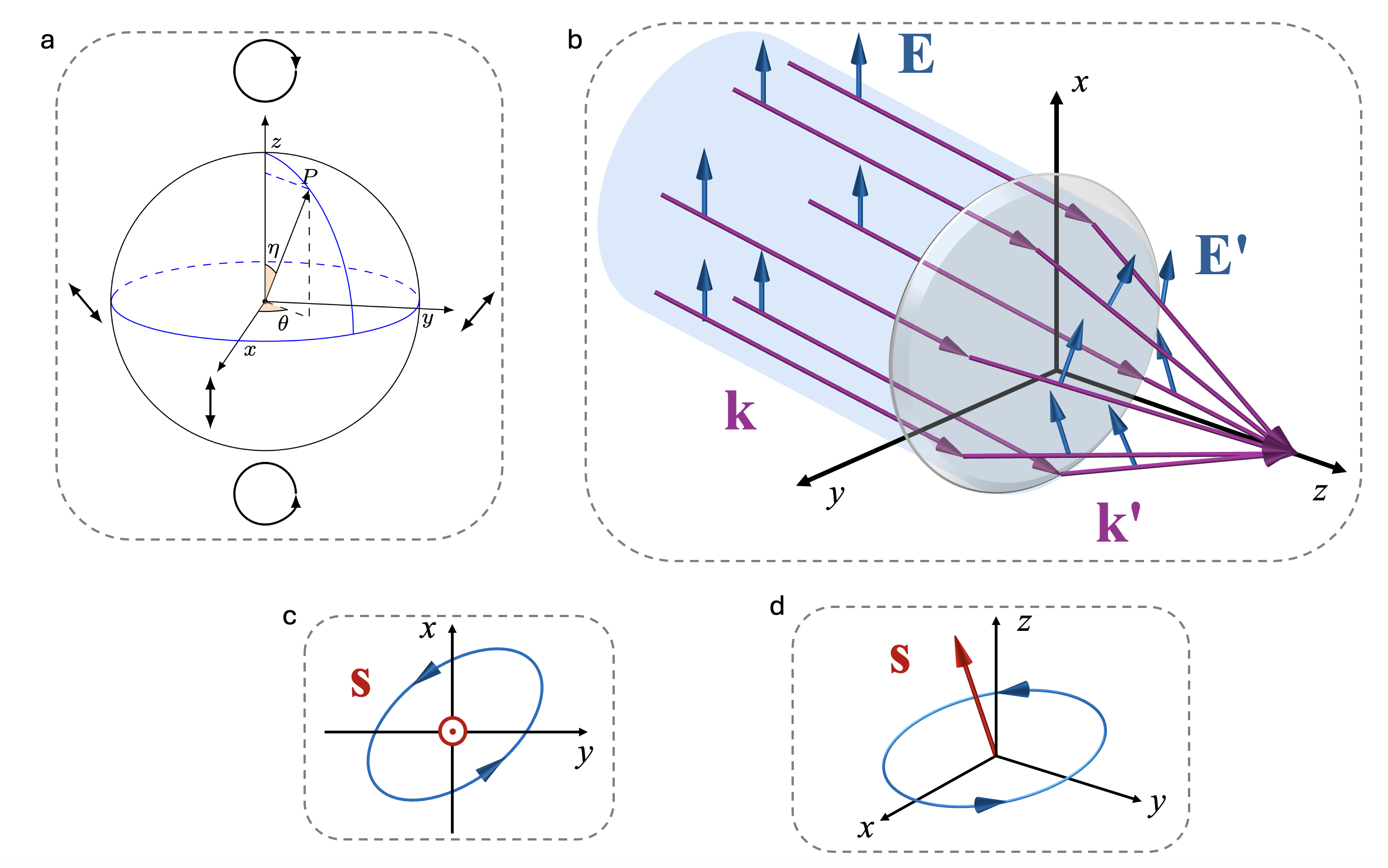}
\caption{ a) Poincar\'e sphere representation of 2D polarisation. Degree of ellipticity given by $\eta$ and orientation by $\theta$. The Poincaré sphere representation places circular polarisation states at the sphere’s poles, with linear polarisation states arranged along the equator; b) before focusing an input paraxial beam each plane-wave component's wave vector $\mathbf{k}$ aligns with the direction of propagation $z$, however focusing generates a conical distribution of $\mathbf{k}^{'}$, and resultant changes in polarisation $\mathbf{E}^{'}$. Before the high NA focal lens the field can be accurately described by $\mathbf{E}^\text{T0}$, at and around the focal point the field becomes highly non-paraxial and is described by $\mathbf{E}^\text{T0+L1+T2}$; c) 2D polarisation ellipse and spin vector $\mathbf{s}$ for a paraxial $z$-propagating field; d) 3D polarisation ellipse and $\mathbf{s}$ for a general non-paraxial field.}
\label{fig:1}
\end{figure}

In the paraxial approximation (Eq.~\eqref{eq:3}), LG modes can have arbitrary 2D polarisation states, which are defined by the Jones vector $[\alpha,\beta]^{\text{T}}$. However, as the beam approaches a focal plane, the polarisation becomes more complex and transitions to a three-dimensional (3D) state~\cite{alonso2023geometric}. In this work, we refer to the polarisation state in the $(x,y)$-plane of the input $z$-propagating paraxial beam, as described by the Jones vector before focusing (i.e., the T0 field components), as the 2D polarisation state. The complete 3D polarisation state of a focused, non-paraxial beam can be derived from Eq.~\eqref{eq:8} by substituting the values of $\alpha$ and $\beta$ that define the 2D polarisation state of the input paraxial beam. For example, $[1,0]^{\text{T}}$ or $[0,1]^{\text{T}}$ correspond to 2D $x$-polarised or $y$-polarised LG beams, respectively, while $[1,\pm i]^{\text{T}}/\sqrt{2}$ represents a 2D circularly polarised (CP) beam. It is important to note that $\alpha$ and $\beta$ can be expressed in terms of the azimuth $\theta$ and ellipticity $\eta$ on the Poincaré sphere, which characterise the orientation and ellipticity of the generalized polarisation ellipse, respectively (see Figure~\ref{fig:1}a). This formalism will be utilised throughout the review.

While incorporating higher-order terms enhances quantitative accuracy, it is crucial for qualitative correctness that \textbf{all} relevant contributions up to a given order are included. For example, when calculating quantities such as the energy density Eq.~\eqref{eq:1} up to second order in the paraxial parameter, it is not sufficient to retain only the `pure' contributions from terms quadratic in T0 and L1. It is equally important to include `interference' terms, such as those involving the product of T0 and T2. As previously noted, paraxial electromagnetic fields in free space exhibit dual symmetry. However, this symmetry is broken in non-paraxial regimes, leading to dual asymmetry in the fields. Consequently, accurate descriptions of optical properties that rely on magnetic contributions necessitate an explicit calculation of the magnetic field. Equally critical is the requirement that both the electric and magnetic fields be derived to the \textit{same order}. Failing to do so can result in optical properties that are incorrectly skewed, either electrically or magnetically biased. A common simplification seen in the literature is to impose \(\nabla \cdot \mathbf{E} = 0\) prior to applying \(\partial \mathbf{B} / \partial t = \nabla \times \mathbf{E}\), while neglecting the remaining Maxwell equations. This can introduce an asymmetry in the T2 field and lead to inaccurate predictions of optical properties. This issue is of particular importance in both analytical and numerical approaches.

\subsection{Vector Vortex Beams} 

The above theory is suitable to describe what are often termed `scalar' beams: beams of light whose local, paraxial polarisation state, is homogeneous across the plane orthogonal to propagation. So-called `vector' beams are modes which possess an inhomogeneous distribution of local polarisation states when paraxial (not to be confused with a `vector description' of beams which is a phrase often used to describe focused scalar beams where the vector nature of light becomes relevant). More generally, vector beams are referred to as vector vortex beams (VVBs) \cite{rosales2018review, arora2020detection}. VVBs are beams which possess both inhomogeneous polarisation structure and azimuthal phase structure. VVBs can be viewed as the superposition of two scalar orthogonally polarised spatial modes (labelled A and B) which individually carry optical OAM through the azimuthal phase factor $\text{e}^{i\ell_{\text{A(B)}}\phi}$. VVBs are also known as higher-order Poincar\'e  (HOP) \cite{milione2011higher} or hybrid-order Poincar\'e  (HyOP) \cite{yi2015hybrid, arora2020detection, ling2016characterization, arora2020hybrid} beams, the former carry the same magnitude of OAM but of opposite sign $\ell_{\text{A}} = -\ell_{\text{B}}$:  vector beams (VBs) \textit{which carry no overall OAM}. The latter carry arbitrary OAM:  vector vortex beams VVBs. The controllable combination of the different degrees of freedom of these significantly structured light beams means they are under intense study due to their applicability in areas such as optical manipulation, optical communications, quantum information, laser machining, and enhanced imaging to name a few \cite{naidoo2016controlled, forbes2019quantum, wang2020vectorial, he2022towards, nape2022revealing, nape2023quantum, bliokh2019geometric, wang2021polarization, zhou2021optical}. The analytical description of VVBs, both paraxial and non-paraxial, is readily provided by using a suitable superposition of Eq.~\eqref{eq:8}:

\begin{align}
    \mathbf{E}^\text{T0+L1+T2}_{\text{VVB}} &=  \mathbf{E}^\text{T0+L1+T2}_{\alpha_\text{A},\beta_\text{A},\ell_\text{A},p_\text{A}} + \mathbf{E}^\text{T0+L1+T2}_{\alpha_\text{B},\beta_\text{B},\ell_\text{B},p_\text{B}}
    \label{eq:10}
\end{align}

where $\bigl(\alpha_\text{A}\hat{\mathbf{x}}+\beta_\text{A}\hat{\mathbf{y}}\bigr)\cdot\bigl(\alpha_\text{B}^*\hat{\mathbf{x}}+\beta_\text{B}^*\hat{\mathbf{y}}\bigr)=0$, i.e. scalar beams A and B are orthogonally polarised. 


\section{Intensity profile} \label{IV}

The energy of an electromagnetic beam is routinely measured through its intensity. As we mentioned above, for a paraxial beam the electric $u_\text{E}$ and magnetic $u_\text{B}$ energy density contributions to the total energy density $u$ are equivalent. For focused, non-paraxial beams, this is not the case and $u_\text{E} \neq u_\text{B}$. Importantly, since most matter is dual-asymmetric – composed of electric charges and not magnetic monopoles – it predominantly interacts with the electric component of the field (e.g. electric-dipole approximation). Consequently, when studying the interaction of light with matter, it is often sufficient to consider only the electric field contribution to the intensity. We therefore look to study the intensity of a focused structured beam using Eq.~\eqref{eq:8}. In order to highlight the most significant influence of non-paraxial fields on the intensity of focused structured light, we provide the analytical expression of the intensity at the focal plane ($z=0$) up to second-order in the paraxial parameter (i.e. calculating $\mathbf{E}^{*\text{T0}}\cdot\mathbf{E}^{\text{T0}}+\mathbf{E}^{*\text{L1}}\cdot\mathbf{E}^{\text{L1}}+\mathbf{E}^{*\text{T2}}\cdot\mathbf{E}^{\text{T0}}+\mathbf{E}^{*\text{T0}}\cdot\mathbf{E}^{\text{T2}}$) with a CP input beam $\alpha=1/\sqrt2, \beta=i\sigma/\sqrt2$ and $p=0$.

\begin{align}
    I_\text{LG}^{'}=\frac{c\epsilon_0}{2}|\mathbf{E}|^2\approx\frac{c\epsilon_0}{2}\Bigl[1+\frac{1}{k^2}\Bigl(\frac{\ell^2-\ell\sigma|\ell|}{r^2}+2\frac{\ell\sigma+|\ell|+2}{w_0^2}-\frac{2r^2}{w_0^4}\Bigr) \Bigr]|\psi^{\ell,0}_{\text{LG}}|^2   \label{eq:11}
\end{align}

The first term in square brackets in Eq.~\eqref{eq:11} arises from the paraxial $\text{T0}$ field contribution, $\mathbf{E}^{*\text{T0}}\cdot\mathbf{E}^\text{T0}$. We refer to this paraxial intensity as $I_{\text{LG}}$ throughout the review. The second term originates from the longitudinal and higher-order transverse field contributions. Notably, setting $\sigma = 0$ does not yield the correct form of Eq.~\eqref{eq:11} for a linearly polarised vortex beam – highlighting the significance of spin effects. The higher-order field contributions depend sensitively on the focusing strength, polarisation helicity $\sigma$ (ellipticity), and wavefront handedness $\ell$. In contrast, the paraxial contribution depends only on the magnitude $|\ell|$ via the LG intensity profile $|\psi^{\ell,p}_{\text{LG}}|^2$. Crucially, the term $\mathbf{E}^{*\text{L1}}\cdot\mathbf{E}^\text{L1}$ is responsible for generating all spin-orbit interactions (SOI), through the $\ell\sigma$ dependence. These terms are invariant under simultaneous inversion of spin and orbital angular momentum ($\ell,\sigma \rightarrow -\ell,-\sigma$), but not under individual flips of either ($\ell,\sigma \rightarrow \ell,-\sigma$ or $-\ell,\sigma$).

For beams with $|\ell| = 1$, the intensity profile $|\psi^{\ell,p}_{\text{LG}}|^2 \propto r^2 e^{-2r^2/w_0^2}$ implies that the term in parentheses, $(\ell^2 - \ell\sigma|\ell|)/r^2$, contributes either zero or a non-vanishing Gaussian. Specifically, for parallel combinations of SAM and OAM ($\ell = \sigma$), the contribution vanishes at $r = 0$, resulting in a central intensity null. Strikingly, for anti-parallel combinations ($\ell = -\sigma$), the beam is not `dark' on-axis – the longitudinal field leads to a nonzero central intensity despite the vortex structure. It is widely appreciated that longitudinal fields contribute to the on-axis intensity for tightly-focused $|\ell|=1$ modes. However, Eq.~\eqref{eq:11} also highlights another spin-orbit term $\propto(\ell\sigma+|\ell|)$ which acts to reduce the intensity contributions with radial dependency $r^{2|\ell|}$ when $\ell=-\sigma$. It is the combination of the two spin-orbit terms, dependent on $\ell$ and $\sigma$, which generate the changing position of the maxima of the intensity distributions for non-paraxial vortices. In Eq.~\eqref{eq:11} we neglected the higher-order $1/(kw)^4$, e.g. $\mathbf{E}^{*\text{T2}}\cdot\mathbf{E}^\text{T2}$ contributions, however these terms also generate SOI and deliver a non-zero on-axis intensity for very tightly focused $|\ell|=2$ beams, though this effect is very weak. 

Ganic et al.~\cite{ganic2003focusing} were the first to predict a non-zero on-axis intensity in the focal plane for tightly-focused scalar vortex beams. Using vector Debye theory they highlighted that for $x$-polarised input vortex beams of charge $\ell=1,2$ there is a non-zero on-axis intensity in the focal plane, however $\ell=3$ modes have an intensity null at $r=0$ even under tight focusing. By plotting the individual contributions $|E_x|^2$, $|E_y|^2$, and $|E_z|^2$, it is clear to see in their numerical simulations that the longitudinal fields $|E_z|^2$ generate the on-axis intensity for $|\ell|=1$ modes (as we highlighted above), whereas for $|\ell|=2$ it stems from the $|E_y|^2$ contribution, which for an input $x$-polarised beam are the second-order transverse fields $\mathbf{E}^{\text{T2}}$. 

Another theoretical study followed soon after \cite{torok2004use} looking at the application of LG beams in STED microscopy. This study predicted similar results for a tightly-focused $x$-polarised beam to Ganic et al.~\cite{ganic2003focusing}, however it also looked at tightly-focused circularly polarised LG beams. Interestingly, because of the demands of STED microscopy the authors only reported the case of when $\ell=\sigma$ as this provides the desired perfect zero on-axis intensity at the focus. 

\begin{figure}[!ht]
    \includegraphics[width = \linewidth]{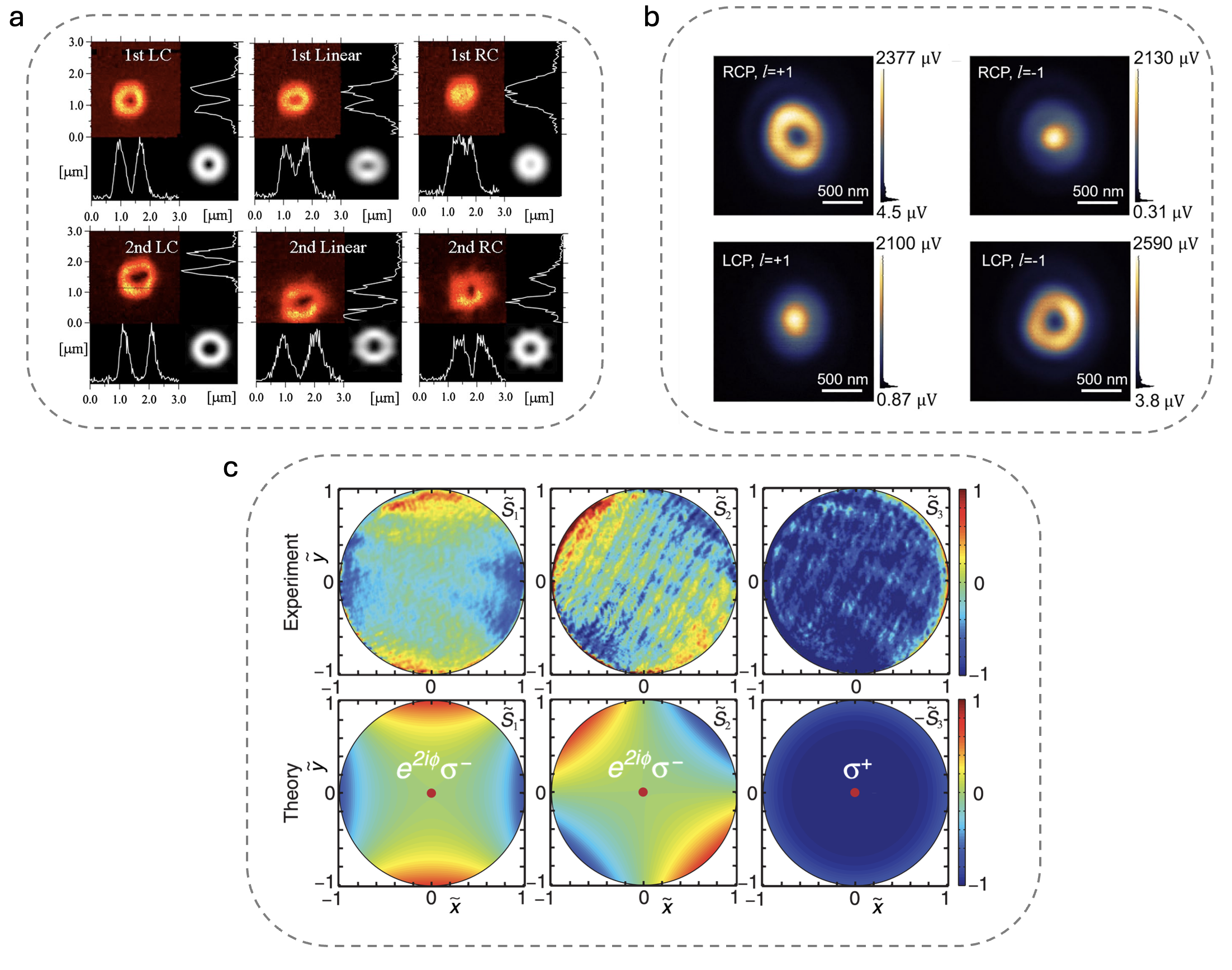}
    \caption {Measurements of intensity and polarisation of focused vortex beams: a) Measured intensity of tightly focused $\ell=1,2$ vortex beams with $\sigma=0,\pm1$ (simulations insets) \cite{iketaki2007investigation}. For $\ell=1 $ the anti-parallel combination $\ell=1,\sigma=-1$ becomes essentially a Gaussian beam due to the longitudinal fields; for the parallel combination  $\ell=1,\sigma=1$ the central intensity is zero; for linearly-polarised $\sigma=0$ beams the central intensity is non-zero and in-between the two $\sigma=\pm1$ in magnitude. For the $\ell=2$ case the on-axis intensity predicted by theory is difficult to see in the experiment due to its very small magnitude (fourth-order with respect to the paraxial parameter; b) Mapped purely transverse $|\mathbf{E}_\text{T}|^2$ (parallel $\sigma,\ell)$  and purely longitudinal $|\mathbf{E}_\text{L}|^2$  (anti-parallel $\sigma,\ell)$ near-field intensity distributions of focused circularly-polarised vortex beams through photoinduced force microscopy (N.B. the opposite sign convention for left- and right-handed $\sigma$ and $\ell$ to what we use in this review) \cite{dong2024nearfield}; c) Distribution of Stokes parameters in the exit pupil of a high-NA optical microscope for a nanoparticle placed precisely at the focal point \cite{rodriguez2010optical}. The scattering of the input circularly-polarised beam showing spin-to-orbit AM conversion as a beam with opposite circular handedness and vortex of charge 2 is generated.}
    \label{fig:2}
\end{figure}

The first experimental proof of the on-axis intensity for tightly focused vortex beams was provided by Bokor et al.~\cite{bokor2005investigation} in 2005. By scanning the focal plane of tightly-focused $x$-polarised and circularly-polarised $\ell=1$ LG modes with a 100nm diameter fluorescent microbead, the resultant fluorescence enabled the mapping of the intensity of the focused beam. As predicted by the preceding theoretical studies, $\ell=\sigma$ beams have a very low central intensity, $\ell=-\sigma$ the central intensity minimum of paraxial vortex beams disappears, and for linearly-polarised beams it is between these two extremes. Two years later the same authors published an experimental study attempting to prove the on-axis intensity and $\ell\sigma$ interplay for $|\ell|=2$ beams~\cite{iketaki2007investigation} – see Figure~\ref{fig:2}a. However, the fact these on-axis intensities for $|\ell|=2$ stem from contributions such as $\mathbf{E}^{*\text{T2}}\cdot\mathbf{E}^\text{T2}$, means that they are proportional to the paraxial parameter to fourth-order $\propto(\lambda/w_0)^4$ and were hard to clearly demonstrate experimentally even for very tightly-focused beams (for $|\ell|=1$ modes the on-axis contributions come from $\mathbf{E}^{*\text{L1}}\cdot\mathbf{E}^\text{L1}$ which is $\propto(\lambda/w_0)^2$ and thus much larger in magnitude).

Using a combination of the helicity basis and Fourier transform techniques, Bliokh et al.~\cite{bliokh2010angular, bliokh2011spin, andrews2012angular} demonstrated that the momentum-space representation of Bessel beams provides unique insight into the intrinsic SOI and intensity structure of non-paraxial vortex beams. In particular, they showed that the $\sigma$-dependent geometric (Berry) phase induces a transverse shift of the intensity maxima.

Dong \textit{et al.}~\cite{dong2024nearfield} inferred the intensity distributions of tightly focused circularly polarised ($\ell=\pm1$) and azimuthally polarised beams using photoinduced force microscopy in the near-field. The time-averaged photoinduced force on the cantilever probe tip is well-approximated by an electric-dipole interaction, consisting of two distinct contributions associated with the transverse electric field components, $|\mathbf{E}_\text{T}|^2$, and the longitudinal component, $|\mathbf{E}_\text{L}|^2$. By measuring these optical forces, they produced force maps that directly revealed the parallel and anti-parallel SOI of tightly focused CP Laguerre–Gaussian beams ($\ell=\pm1$) in the near-field – see Figure~\ref{fig:2}b.

Using a non-mechanical method based on a lens–scatterer–lens configuration, Bliokh et al.~\cite{bliokh2011spin, rodriguez2010optical} demonstrated that fine, subwavelength spin-to-orbit AM conversion can be translated into the paraxial far field. Specifically, they showed that when a tightly focused CP beam is scattered and subsequently recollimated by a high NA lens, the resulting paraxial field exhibits the transformation:
\begin{align}\label{eq:200}
    \ket{\ell,\sigma} \rightarrow a\ket{\ell,\sigma} - b\ket{\ell \pm 2,\sigma \mp 1},
\end{align}
where $a$ and $b$ are coefficients determined by the degree of non-paraxiality. This superposition leads to spatially inhomogeneous polarisation in the output beam, manifesting as a characteristic four-lobed pattern in the first and second Stokes parameters~\cite{brasselet2009optical} – see Figure~\ref{fig:2}c.

An application of the $\ell$- and $\sigma$-dependent intensity structure of a focused vortex beam was explored in the context of a form of dichroism in nano-plasmonic apertures~\cite{zambrana2014angular}. Interestingly, the more Gaussian-like anti-parallel combinations exhibited lower transmission through the apertures compared to the parallel class, whose ring-shaped intensity profile coupled more efficiently. This phenomenon, termed \textit{angular momentum dichroism}, does not represent a true chiral light–matter interaction; rather, it arises from a differential response of the material to beams with distinct spatial intensity distributions.

\begin{figure}[!ht]
    \includegraphics[width = \linewidth]{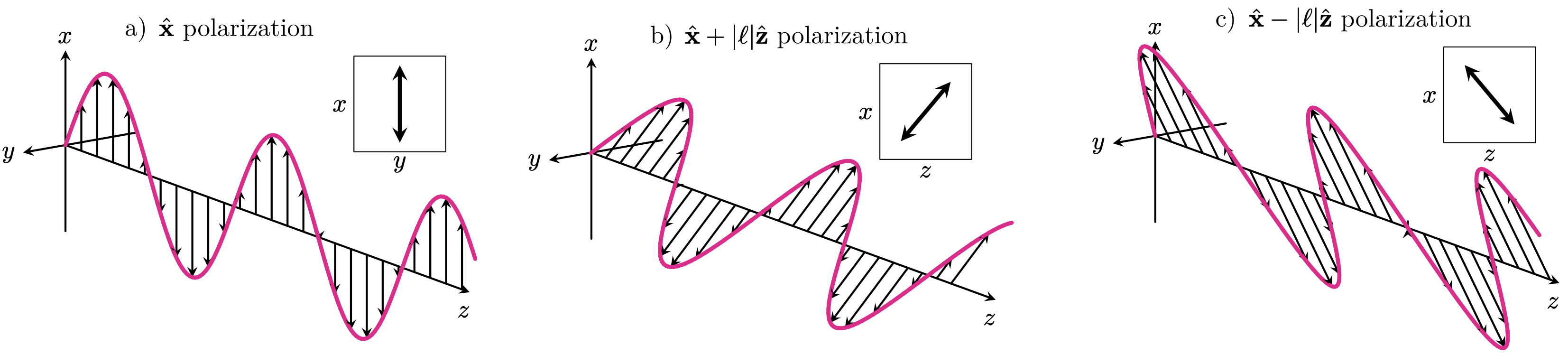}
    \caption{ States of linear polarisation for focused a $z$-propagating beam: a) paraxial 2D $x$-polarised; b) 2D x-polarised optical vortex light with an in-phase longitudinal field component for positive values of $|\ell|$. The polarisation vector is tilted in the positive z-direction in the $xz$-plane; c) same as b) but for negative values of $|\ell|$. In this case, the polarisation vector is tilted toward the negative $z$-direction.}
    \label{fig:3}
\end{figure}

Inspection of Eq.~\eqref{eq:8} reveals that for beams with $\ell = 0$, the T0 and L1 field components are always $\pi/2$ out of phase with one another – a key feature of transverse spinning light, as discussed in Section~\ref{VI}. However, when $\ell \neq 0$, in-phase contributions between the T0 and L1 components emerge, with their relative phase depending explicitly on $\ell$. This $\ell$-dependence plays a significant role in shaping the polarisation structure in the focal plane, where the input electric field in the $xy$-plane becomes tilted into or out of the $z$-direction depending on the sign of $\ell$ (Figure~\ref{fig:3}). In anisotropic materials, this effect gives rise to a topological-charge-dependent dichroism and birefringence, analogous to the conventional linear dichroism and birefringence observed under unstructured illumination~\cite{forbes2024topological}. 

In anisotropic epsilon-near-zero (ENZ) metamaterials, Aita et al.~\cite{aita2025longitudinal} have shown how  the longitudinal field component of vector beams plays a crucial role, particularly for radially polarised beams, where it is required to maintain the divergence-free nature of the electric field. The anisotropic absorption in such metamaterials redistributes energy between the longitudinal and transverse components, enabling control over beam propagation, mode content, and spatial polarisation distributions – especially prominent in the ENZ regime. Notably, they showed how under \textit{low} NA focusing, the longitudinal field propagating in the ENZ material can exceed the transverse field in strength.

More generally Eq.~\eqref{eq:8} reveals that the full three-dimensional polarisation ellipse of a focused vortex beam is highly intricate near the focal plane. As discussed previously for linearly polarised light, to first order in the paraxial parameter, the longitudinal field of a vortex beam introduces both an in-phase tilting in the propagation direction (Figure~\ref{fig:3}) and a $\pi/2$ out-of-phase spinning (degree of ellipticity) in the transverse plane (Figure~\ref{fig:6}). Furthermore, the spatial distribution of the longitudinal component causes the three-dimensional polarisation ellipse to vary across the transverse beam profile and along the propagation direction. For instance, to first order in $1/kw$, the full three-dimensional polarisation ellipse of an $x$-linearly polarised beam takes the form $\mathbf{\hat{x}}+\mathbf{\hat{z}}\frac{1}{k}((i[\frac{|\ell|}{r}-\frac{2r}{w^2}]+\frac{kr}{R[z]})\cos\phi+\frac{\ell}{r}\sin\phi)$
showing an acute dependence on $r$, $\phi$, and $z$,  as well as both the sign and magnitude of $\ell$. Ouyang~\textit{et al.}~\cite{ouyang2021synthetic} exploited the controllable three-dimensional polarisation structure of tightly focused vortex beams to demonstrate dichroism (differential absorption) of optical vortices in anisotropic nanoparticles – an effect they termed \textit{synthetic helical dichroism}. They showed that the distinct absorption responses of achiral nanoparticles to different tightly focused vortex beams enabled six-dimensional optical information multiplexing in self-assembled plasmonic nanostructures. Moreover, synthetic helical dichroism has been extended to so-called perfect optical vortices~\cite{pinnell2019perfect,yang2022ultra}.

It is well established that the longitudinal electric fields of radially polarised HOP beams (with zero OAM) can achieve significantly smaller on-axis focal spots than conventional scalar beams~\cite{zhan2009cylindrical}. Building on this, Hao et al.~\cite{hao2010phase} were the first to demonstrate that azimuthally polarised HyOP beams (carrying non-zero OAM) can also produce highly confined focal spots – remarkably, through purely transverse electric fields (interestingly this result is hinted at in a preceding study~\cite{sato2008hollow}). This counterintuitive result was experimentally confirmed by Li et al.~\cite{li2014super}, who achieved a super-resolved focal spot with a 31\% reduction in lateral focal area at NA$=1.4$ compared to CP scalar beams. An excellent overview on producing small bright and dark focal spots with vector beams can be found in Ref.~\cite{kozawa2021small}.   



\section{Poynting, Orbital, and Spin momentum} \label{V}

Electromagnetic momentum density $\mathbf{p}$ is conventionally expressed via the (kinetic) Poynting vector, $\mathbf{\Pi} = c^2\mathbf{p}= \frac{c^2\epsilon_0}{2}\text{Re}[\mathbf{E}^* \times \mathbf{B}]$, which represents the energy \textit{flow} density. It is now well-established that, for monochromatic fields, $\mathbf{p}$ can be decomposed into two physically distinct components $\mathbf{p} = \mathbf{p}_{\text{o}} + \mathbf{p}_{\text{s}}$ : the spin momentum density $\mathbf{p}_{\text{s}}$ and the (canonical) orbital momentum density $\mathbf{p}_{\text{o}}$, which correspond to the energy flows associated with the spin and spatial degrees of freedom, respectively~\cite{bekshaev2007transverse, berry2009optical}.       

There has been considerable interest in the individual contributions of spin and orbital momentum densities, e.g. see review~\cite{bekshaev2011internal}, particularly in how they influence mechanical forces on particles. Strikingly, in light-matter interactions, it is generally \textit{not} the kinetic momentum $\mathbf{p}$ that governs the opto-mechanical action on particles. Rather, the situation is more nuanced and highly dependent on the particle’s size and intrinsic material properties. For Rayleigh particles ($ka \ll 1$) with electric polarizability $\alpha_e$, the dominant electric dipole forces include a gradient force arising from intensity variations $\mathbf{F}_\text{Gradient}\propto\text{Re}(\alpha_\text{e})\nabla I_\text{e}$ and a radiation pressure force that is proportional to the canonical momentum $\mathbf{p}_{\text{o}}$, $\mathbf{F}_\text{Pressure}\propto\text{Im}(\alpha_\text{e})\mathbf{p}_{\text{o}}$ – \textit{not} the Poynting momentum \cite{bliokh2014extraordinary}. This is why $\mathbf{p}_{\text{s}}$ has been described as a ``virtual'' quantity: it does not appear to produce observable effects in conventional light-matter interactions~\cite{bliokh2013dual}.

The formula for calculating the (electric) canonical momentum density is given in Box 1, which highlights that the azimuthal component of the canonical momentum density $p_{\mathbf{E}}^{\text{o}, \phi}$ up to second-order in the paraxial parameter can be calculated as follows:
\begin{align}\label{eq:18}
     p_{\mathbf{E}}^{\text{o}, \phi} &= \frac{\varepsilon_0}{2 \omega r}\,\text{Im}\Bigl( \mathbf{E}^{*\text{T0}}\frac{\partial}{\partial \phi}\mathbf{E}^{\text{T0}} + \mathbf{E}^{*\text{L1}}\cdot\frac{\partial}{\partial \phi}\mathbf{E}^{\text{L1}}+ \mathbf{E}^{*\text{T2}}\cdot\frac{\partial}{\partial \phi}\mathbf{E}^{\text{T0}}+ \mathbf{E}^{*\text{T0}}\cdot\frac{\partial}{\partial \phi}\mathbf{E}^{\text{T2}}\Bigr).
\end{align}
We concentrate on the azimuthal component (over the universal longitudinal component) as it is responsible for the orbital angular momentum $l_z=rp^{\text{o},{\phi}}$ and causes circulation of probe particles. Carrying out the calculation Eq.~\eqref{eq:18} using Eq.~\eqref{eq:8} with $\alpha=1/\sqrt2$, $\beta=i\sigma/\sqrt2$, i.e. a circularly polarised LG beam gives:
\begin{align}
    p_{\mathbf{E}}^{\text{o}, \phi} &= \frac{I_{\text{LG}}}{c\omega} \Biggl[ \frac{\ell}{r} 
    + \frac{(\ell+\sigma)}{2k^2r} \left( |\gamma|^2 - 2\,\text{Re}\left(\frac{\ell\sigma}{r} \gamma\right) + \frac{\ell^2}{r^2} \right) 
    - \frac{\ell}{k^2r} \text{Re} \left( \gamma^2 + \gamma' + \frac{\gamma}{r} - \frac{\ell^2}{r^2} \right) \Biggr].\label{eq:19}
\end{align}
The first term in square brackets represents the \textit{paraxial contribution} from T0. The second group of terms, proportional to $\ell + \sigma$, originates from L1. The final group, proportional to $\ell/k^2$, arises from the interference between T0 and T2.

This expression highlights that all SOIs arises solely from the \textit{longitudinal field L1}. The higher-order transverse correction (T2) modifies the paraxial canonical momentum density but does not introduce SOI on its own (to this order of approximation). As expected, for CP Gaussian beams where $\ell = 0$, both the paraxial and the T2 correction terms vanish – confirming that SOI is entirely due to longitudinal contributions at this level. The most striking and immediate result is that while a paraxial Gaussian beam with $\ell=0$ has zero azimuthal canonical momentum density (and thus zero OAM), under tight focusing a CP Gaussian mode has a non-zero azimuthal orbital momentum density and OAM: this is known as spin-to-orbit AM conversion \cite{bliokh2011spin, bliokh2015spin}.

An important point to note at this juncture is that, although the phenomenon is framed in terms of spin-to-orbit AM conversion, experiments typically measure the local forces on probe particles smaller than the beam waist. Consequently, these experiments effectively probe the azimuthal component of the canonical momentum density (Eq.~\eqref{eq:19}), from which the OAM can be inferred $\mathbf{l}= \mathbf{r}\times\mathbf{p}^\text{o}$. This is because probe particles – being smaller than the beam waist – circulate around the beam due to the azimuthal component \(\phi\) of the linear momentum density. While the well-known OAM of vortex beams is directed along the propagation axis \(z\) and can induce spinning of particles about their own axis~\cite{padgett2011tweezers}, it does not cause particles to orbit the beam axis. This subtle distinction is frequently overlooked in discussions of OAM, where “OAM” often actually refers to the azimuthal component of the linear momentum density. Indeed, if experiments were to measure the integrated AM values (i.e. particles larger than the beam width), the familiar torque from the SAM of a circularly polarised Gaussian beam would be indistinguishable from any additional torque produced by the SOI-generated OAM.

\begin{figure}[!ht]
    \includegraphics[width = \linewidth]{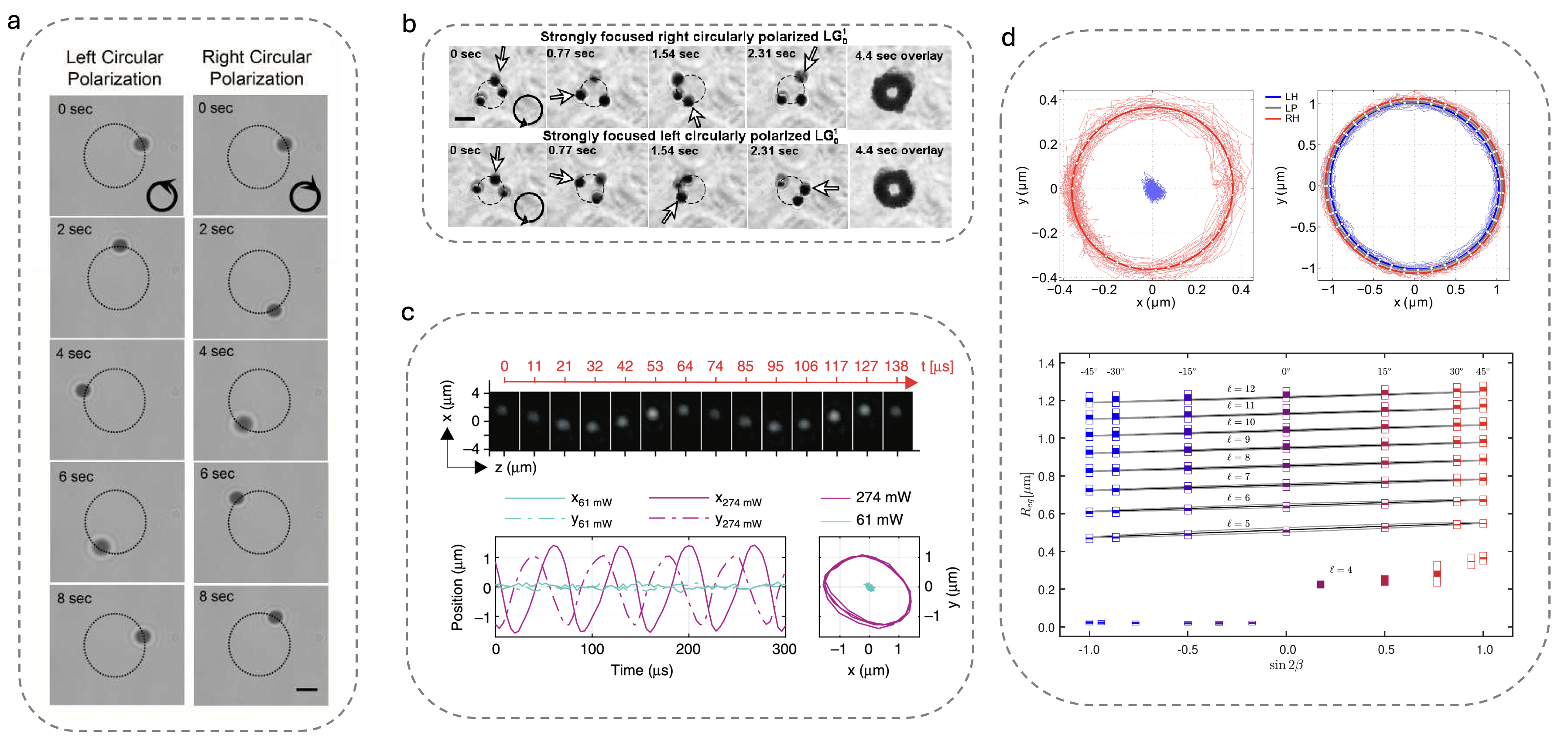}
    \caption{a) orbital motion of 3$\mu$m spherical gold particle in a tightly-focused Gaussian beam~\cite{zhao2009direct}; b) successive frames of a video showing the orbital motion of three gold particles 3.0-5.5$\mu$m in diameter trapped in a tightly-focused vortex beam~\cite{zhao2007spin}; c) orbital motion of dieletric microspheres due to the spin momentum in a paraxial circularly polarised Gaussian vacuum trap~\cite{svak2018transverse}; d) orbital motion of polystyrene microspheres trapped in tightly focused circularly polarised LG beams (N.B $\sin2\beta$ is the degree of polarisation ellipticity)~\cite{arzola2019spin}.}
    \label{fig:4}
\end{figure}

Originally predicted in 2004 \cite{nieminen2008angular}, the first claim of observing the spin-to-orbit AM conversion came in 2007 \cite{zhao2007spin}, where tightly focused circularly polarised LG modes with $\ell = -1$, $p = 0$ were shown to induce a $\sigma$-dependent rotational speed of three gold micro-particles trapped in the high-intensity ring of the beam through the radiation pressure force $\mathbf{F} \propto \mathrm{Im}(\alpha) \mathbf{p}^{\mathrm{o}}$ – see Figure~\ref{fig:4}b. For $\sigma = \ell = -1$ (parallel SAM–OAM configuration), the circulation of the gold particles was $22.5 \pm 5.2\%$ faster than for $\sigma = 1, \ell = -1$ (antiparallel).

Setting $\ell=0$, Eq.~\eqref{eq:19} shows that CP Gaussian beams also undergo spin-to-orbit AM conversion. However, experimentally verifying this effect is challenging because it originates from the longitudinal fields, which in Gaussian modes are generally much weaker than the transverse components even under tight-focusing. Furthermore, these longitudinal fields produce an intensity distribution proportional to \(r^2\), however particles are predominantly trapped in the on-axis high-intensity region of the beam rather than in the outer ring where the OAM manifests. To circumvent this, Zhao et al.~\cite{zhao2009direct} employed a counterintuitive approach by using a more weakly focused CP Gaussian beam, which had longitudinal fields at only about 10\% of the transverse fields (Figure~\ref{fig:4}a). Although weaker, this configuration reduced the on-axis trapping efficiency sufficiently to allow micron-sized gold particles to orbit, with the direction of rotation reversing upon switching the handedness $\sigma$ of the input fundamental Gaussian beam. Similar orbital motion was earlier inferred from the diffraction patterns of tightly focused CP Gaussian beams~\cite{adachi2007orbital}.

In a comprehensive experimental and theoretical study, Arzola et al.~\cite{arzola2019spin} investigated the orbital dynamics of polystyrene microspheres trapped in circularly polarised LG beams with topological charges up to \(\ell = 12\), tightly focused using a high NA (1.2) lens (Figure~\ref{fig:4}d). The equilibrium orbiting radii of the particles were found to depend strongly on the ellipticity of the input beam, with smaller radii observed for anti-parallel combinations of SAM and OAM. Notably, when SAM and OAM were anti-parallel (\(\mathrm{sgn}(\ell) \neq \mathrm{sgn}(\sigma)\)), particles were trapped at the beam centre and ceased to circulate altogether for \(\ell \leq 4\). Interestingly, the orbital frequency was higher for anti-parallel SAM-OAM configurations, contrasting with the results of Zhao \textit{et al.}~\cite{zhao2007spin} – an effect likely due to stronger radial confinement in beams with higher \(\ell\).

A theoretical study of the temporal evolution of spin-to-orbit angular momentum conversion in tightly focused, circularly polarised Gaussian pulses demonstrated a strong sensitivity to both pulse duration and central wavelength~\cite{zhang2023temporal}. Further mechanical observations of spin-to-orbit AM conversion have been reported~\cite{huang2021spin, wang2010optical}, although it is important to note that the theoretical interpretations presented in those studies have since been shown to be incorrect~\cite{yan2011comment, bekshaev2011comment} and lead us into returning to the spin momentum density $\mathbf{p}_\text{s}$. 

Box 1 shows that, \textit{even in the paraxial regime}, the spin momentum is directed azimuthally and its magnitude is comparable to the azimuthal orbital momentum density. For a \(p=0\) beam, the azimuthal Poynting momentum density is explicitly given by
\begin{align}\label{eq:spin}
    p_{\phi}=p^{\mathrm{o}}_{\phi} + p^{\mathrm{s}}_{\phi} &= \frac{I_{\mathrm{LG}}}{c \omega} \left[\frac{\ell}{r} + \sigma \left(\frac{2r}{w^2} - \frac{|\ell|}{r} \right) \right].
\end{align}
Thus, if $\mathbf{p}$ can be coupled through an optical force, and hence \(p_{{\phi}}^{\mathrm{s}}\), orbital motion may be observed regardless of any spin-to-orbit AM conversion, even without tight focusing and within the paraxial regime. In fact, this force should be significantly stronger than that arising from SOI effects due to its origins in the paraxial (T0) fields. Indeed, the orbital motion of particles in paraxial CP fields – where longitudinal fields and spin-to-orbit AM conversion are effectively negligible – due to spin momentum has been experimentally observed multiple times \cite{angelsky2012orbital, ruffner2012optical, svak2018transverse}.

Studies~\cite{bekshaev2013subwavelength, bliokh2014extraordinary, hayat2015lateral, antognozzi2016direct} have demonstrated that for magnetoelectric dipolar particles, the spin flow can induce forces through dipole-dipole interference effects, with \(\mathbf{F} \propto -k^3 \mathrm{Re}(\alpha_{\mathrm{e}} \alpha_{\mathrm{m}}) (\mathbf{p}^{\mathrm{o}} + \mathbf{p}^{\mathrm{s}})\). More generally, such coupling can occur via higher-order multipoles (quadrupolar, octupolar, etc.) in non-magnetic (non-electric) particles as well.

\begin{figure}[!ht]
    \includegraphics[width = \linewidth]{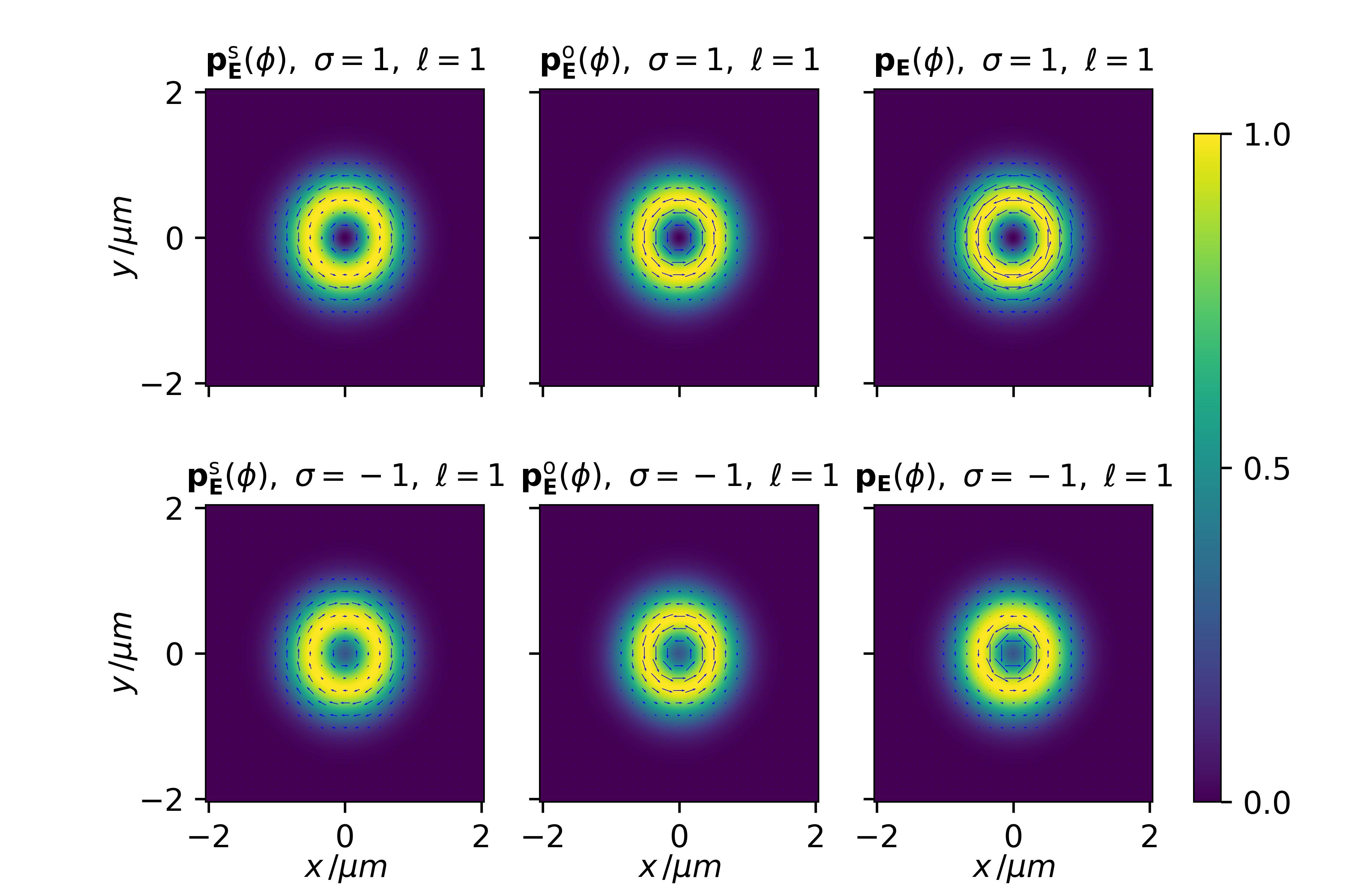}
   \caption{Azimuthal components of the spin momentum (first column), canonical (orbital) momentum from Eq.~\eqref{eq:14} (second column), and total (Poynting) momentum (third column) densities for focused, circularly polarised Laguerre–Gaussian beams with $\ell=1$, $p=0$, and $\lambda = w_0$, overlaid on the intensity distribution from Eq.~\eqref{eq:11}. It is evident that the spin momentum distribution is significantly more sensitive to the sign of $\sigma$ than the canonical momentum density, which is responsible for spin-to-orbit angular momentum conversion. Moreover, the spin momentum strongly influences the behaviour of the total (Poynting) momentum. For instance, a micro-sized particle engaging with the full Poynting momentum in the high-intensity region would experience a marked change in its orbital frequency upon flipping the sign of $\sigma$, an effect predominantly governed by the spin momentum rather than the orbital component. (All plots are normalised to the top-left panel.)}
    \label{fig:5}
\end{figure}

The spin flow up to second-order can be readily calculated using the theoretical formalism used throughout this report. Unlike the other quantities in this review which we give explicitly, we choose not to with this quantity because the analytical simplicity is lost in cases where $p\neq0$ due to the radial gradient of the square of Laguerre polynomial. Nonetheless, the spin, orbital, and total Poynting momentum densities for non-paraxial LG modes are plotted in Figure~\ref{fig:5}. It is clear to see that $\mathbf{p_\text{s}}$ is much more sensitive to the input beam values of $\sigma$ and $\ell$ and determines the behaviour of the full Poynting momentum $\mathbf{p}$. 

In light of this, all published studies claiming mechanical observation of spin-to-AM conversion to date involve particles well within the Mie regime, which, as discussed, generally experience forces proportional to the full momentum density $\mathbf{p}$. While spin-to-AM conversion (Eq.~\eqref{eq:19}) undoubtedly contributes, the observed effects cannot be attributed solely to it (numerical simulations may enable further insight into the partition). Furthermore, spin-to-AM conversion originates from non-paraxial field components and, as a field quantity in isolation, is significantly weaker than the azimuthal component of the spin momentum in a paraxial beam. It is important to emphasise that particle properties critically influence both the magnitude and sign of the resulting forces. Given the complexity and multiplicity of multipolar contributions to the azimuthal forces in these systems, fully disentangling their effects is essentially impossible. Therefore, an unequivocal mechanical demonstration of spin-to-orbit AM conversion requires observing orbiting motion of a Rayleigh particle (forces solely due to $\mathbf{p}_\text{o}$) induced by a circularly polarised Gaussian beam: a significant challenge for the future.

\section{Angular momentum} \label{VI}

Light carries both spin and orbital angular momentum (SAM and OAM). SAM is intrinsic and associated with polarisation ellipticity in a given transverse plane, while OAM can be intrinsic or extrinsic. This review focuses on intrinsic OAM in vortex beams, arising from the beam’s azimuthal phase gradient. Sadowsky in 1899~\cite{mcdonald2009orbital} and Poynting in 1909~\cite{poynting1909wave} linked circular polarisation with SAM. Beth experimentally confirmed its existence in 1936 by measuring the torque on a birefringent plate~\cite{beth1936mechanical}. As discussed, in 1992, Allen et al.~\cite{allen1992orbital} showed that LG modes carry intrinsic OAM of $\ell \hbar$ per photon. Experimental confirmation came in 1995 when He et al.~\cite{he1995direct} demonstrated that an absorptive microscopic particle trapped in an LG beam rotates due to angular momentum transfer, with the torque reflecting the beam’s OAM. This provided direct evidence of the mechanical effects of optical OAM.

In the paraxial regime, spin and orbital angular momenta are well-defined, with each photon carrying $\sigma \hbar$ of SAM and $\ell \hbar$ of OAM (Box 1). In free space, paraxial beams exhibit no coupling between SAM and OAM, and both quantities are conserved during propagation. However, in the non-paraxial regime – such as in tightly focused beams – this separation breaks down~\cite{barnett1994orbital, bliokh2010angular, bliokh2014conservation}. Under these conditions, the angular momentum properties of light exhibit striking departures from their paraxial counterparts, leading to rich and often surprising behavior. We now turn our attention to these remarkable properties and the physical insights they reveal.

\subsection{Spin Angular Momentum} 

The electric SAM density is given as 

\begin{align}
\mathbf{{s}}_{\text{E}}=\frac{\epsilon_0}{2\omega}\text{Im}(\mathbf{E^*\times\mathbf{E}}).
\label{eq:12}
\end{align}

\begin{figure}[!ht]
    \includegraphics[width = \linewidth]{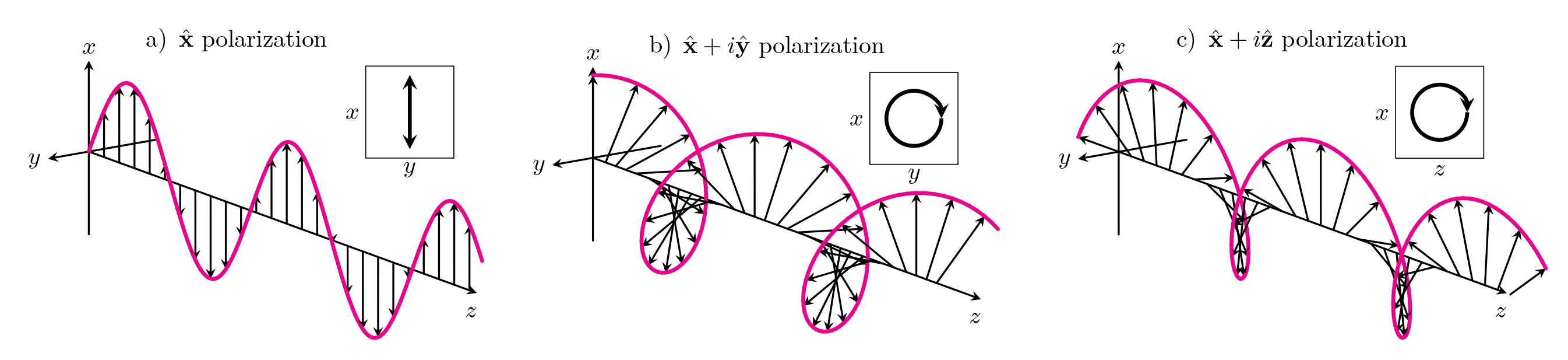}
    \caption{a) Linearly polarised light which has no sense of rotation in either time or space and thus no spin; b) Circularly-polarised light in the plane transverse to propagation generates a \textit{helicoidal} structure on propagation and produces spin AM directed along $\hat{\mathbf{z}}$; c) The $\pi/2$-out-of-phase longitudinal electric field component produces circular polarisation in-plane, thus generating \textit{cycloidal} trajectory and a spin AM out-of-plane (orthogonal to  $\hat{\mathbf{z}}$).}
    \label{fig:6}
\end{figure}

Eq.~\eqref{eq:12} tell us very simply that a generation of SAM density in any given direction requires two orthogonal field components to be $\pi/2$ out of phase with one another, i.e. they must possess a degree of ellipticity. The most well-known example is a paraxial $z$-propagating beam generating a spin density of $\mathbf{s}=\mathbf{\hat{z}}I\sin2\eta/c\omega$, where $\eta$ is the degree of ellipticity. This spin is referred to as longitudinal (L-SAM) because it points in the same direction as propagation, note that it is zero for linearly polarised beams $\eta=0$ and maximum for circularly polarised beams $\eta=\pm\pi/4$. However, it is readily apparent that a longitudinal component to the electric field affords the opportunity for the electric field vector to spin orthogonally to the direction of propagation: transverse spin angular momentum (T-SAM). Thus we may decompose SAM into L-SAM which possess helicoidal trajectories in space and T-SAM with cycloidal trajectories in space (see Figure~\ref{fig:6}). Working up to second-order in the paraxial parameter, the relevant contributions to the spin Eq.~\eqref{eq:12} will be $\mathbf{T_0}\times \mathbf{T_0}$, $\mathbf{T_0}\times \mathbf{L_1}$, and $\mathbf{T_0}\times \mathbf{T_2}$. For input beams with a degree of ellipticity in their 2D state of polarisation, the dominant contribution to the spin is the paraxial contribution $\mathbf{T_0}\times \mathbf{T_0}$ which produces the well known result $\mathbf{s}=\mathbf{\hat{k}}I\sin2\eta/c\omega$ mentioned above. The $\mathbf{T_0}\times \mathbf{L_1}$ term is the leading order contribution to the extraordinary T-SAM, whilst the $\mathbf{T_0}\times \mathbf{T_2}$ contribution generates L-SAM under highly novel conditions: we will give each their own respective subsection below to go through their properties in detail. 

\subsubsection{Transverse Spinning Light} 

Transverse spin, or T-SAM, was first described in 2012 by Bliokh et al.~\cite{bliokh2012transverse} and Kim et al.~\cite{kim2012time} for evanescent waves (a spatially confined electromagnetic field). A remarkable application of transverse spin in evanescent waves is its ability to provide spin-controlled unidirectional propagation of light~\cite{bliokh2015transverse, bliokh2015quantum, bliokh2015spin, lodahl2017chiral}. It is interesting to reflect on the fact that T-SAM is now recognised as a universal feature of spatially confined electromagnetic fields (see reviews~\cite{aiello2015transverse, shi2023advances}), yet the underlying field structures were already present in classical optics. For instance, Richards and Wolf showed in their 1959 paper~\cite{richards1959} that a longitudinal electric field component arises in tightly focused beams, and is $\pi/2$ out of phase with the transverse field — a hallmark of transverse spin that remained conceptually underappreciated for decades. Moreover, there are a wealth of studies describing and measuring the polarisation properties of evanescent waves and tightly-focused beams, but none associated the phase relationship with spin angular momentum~\cite{bliokh2015transverse}. 

\begin{figure}[!ht]
    \includegraphics[width = \linewidth]{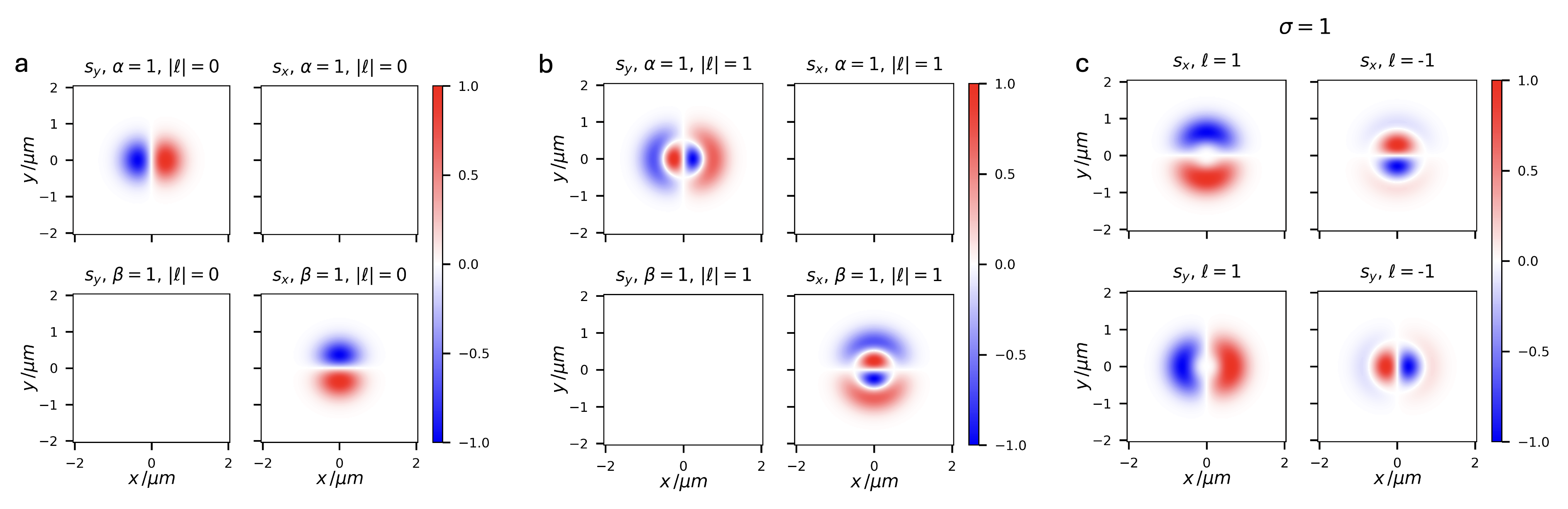}
    \caption{Simulations of transverse spin momentum density for different values of input beam parameters using Eq.~\eqref{eq:13}. a) 2D linearly polarised Gaussian mode; b) 2D linearly polarised LG beam; c) 2D left-circularly polarised LG beam. (In all plots $\lambda=w_0$ and each are individually normalised.)}
    \label{fig:7}
\end{figure}

\begin{figure}[!ht]
    \includegraphics[width = \linewidth]{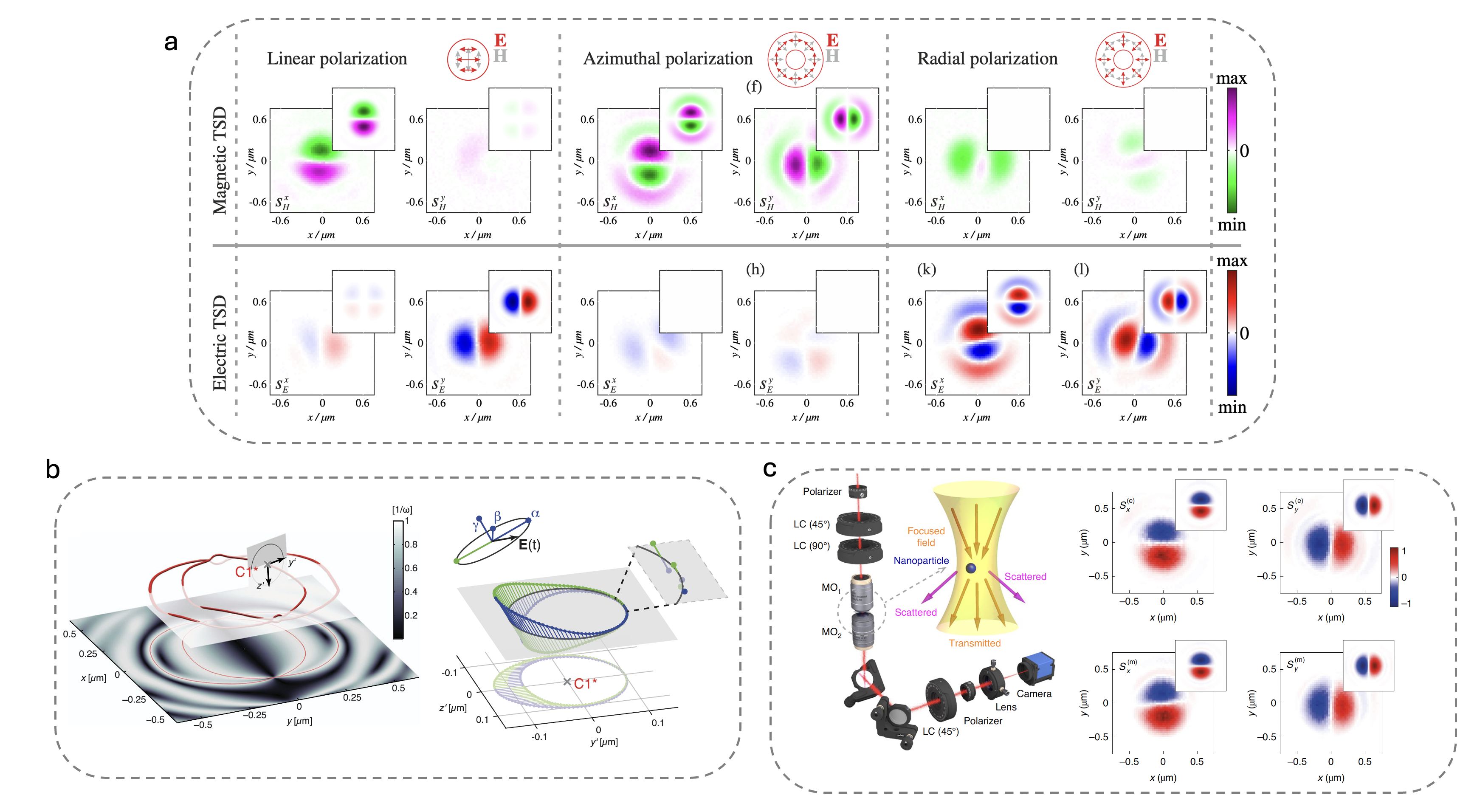}
    \caption{Experimental observations of transverse spin momentum of focused beams. a) Electric and magnetic transverse spin densities for tightly focused modes with various states of 2D linear polarisation \cite{neugebauer2018magnetic} (N.B. the extremely weak $x$-component of the electric T-SAM for the $x$-polarised Gaussian mode is the $\mathbf{L_1}\times \mathbf{T_2}$ contribution mentioned in the text); b) M{\"o}bius strips of light \cite{bauer2016optical}; c) Electric and magnetic T-SAM of unpolarised tightly-focused Gaussian beams \cite{eismann2021transverse} (N.B. the electric and magnetic contributions are identical).}
    \label{fig:8}
\end{figure}

The transverse spin density for a LG mode is calculated using Eq.~\eqref{eq:8} (truncated to first-order in the paraxial parameter, i.e. including $\text{T}0$ and $\text{L}1$, but neglecting $\text{T}2$) into Eq.~\eqref{eq:12} giving
\begin{align}
\mathbf{s_E}^{\text{T0} \times \text{L1}} &=\frac{I_{\text{LG}}}{c\omega}\text{Re}\Bigl\{\frac{2}{k}\Bigl[ \mathbf{\hat{x}} \Bigl(\alpha\beta^* \bigl(\gamma\cos\phi-\frac{i\ell}{r}\sin\phi\bigr) + |\beta|^2\bigl(\gamma\sin\phi+\frac{i\ell}{r}\cos\phi\bigr)\}\Bigr) \nonumber \\ 
        & - \mathbf{\hat{y}} \Bigl(\alpha^{*}\beta \bigl(\gamma\sin\phi+\frac{i\ell}{r}\cos\phi\bigr) + |\alpha|^2\bigl(\gamma\cos\phi-\frac{i\ell}{r}\sin\phi\bigr)\}\Bigr) \Bigr]\Bigr\}. \label{eq:13}
\end{align}
(Note that $\mathbf{L_1}\times \mathbf{T_2}$ generates transverse spin but is extremely weak even under very tight focusing, e.g. see Figure~\ref{fig:8}a). The spatial distributions of Eq.~\eqref{eq:13} can be found in Figure~\ref{fig:7} for various input beam parameters. There are two important limiting cases of Eq.~\eqref{eq:13}: $\alpha,\beta$ are real, which corresponds to input 2D linearly polarised beams; and when $\alpha$ is real but $\beta$ is imaginary, which corresponds to input 2D elliptically polarised beams. Concentrating on the case where $\alpha,\beta$ are real – as this allows the disentanglement of the L-SAM associated with polarisation ellipticity in the $xy$-plane – gives the transverse spin momentum density in terms of the Poincar\'e angle $\theta$:
\begin{align}
\mathbf{s_E}^{\text{T0} \times \text{L1}}_\text{Lin} &=\frac{I_{\text{LG}}}{c\omega}\frac{\text{Re}}{k}\bigl[ \mathbf{\hat{x}}\gamma(\sin^2\theta\sin\phi-\sin2\theta\cos\phi)  
         - \mathbf{\hat{y}}\gamma (\cos^2\theta\cos\phi-\sin2\theta\sin\phi)\bigr]. \label{eq:14}
\end{align}
In stark contrast to paraxial L-SAM, immediately we see that the spin is directed transverse to the direction of propagation $z$, is independent of helicity $\sigma$, and is dependent on the orientation of the state of input linear polarisation.  The integral value of transverse spin for focused beams is evidently zero $\int\mathbf{s}^{\text{T0} \times \text{L1}} d\mathbf{r}_\perp=0$ (A 3D integration over all space diverges for beams; therefore, transverse 2D integrals over the beam cross-section are used, yielding densities per unit length \cite{bliokh2015transverse}).

Evidence that tightly-focused beams could generate transverse \textit{angular momentum} was first theoretically predicted and inferred experimentally in 2013~\cite{refId0}. In their experiment, Banzer et al. indirectly highlighted the existence of the `photonic wheel' (purely transverse AM) by measuring the intensity at the focal plane of a specifically designed Hermite-Gauss mode (TEM10) where one lobe is left-circularly polarised and the other lobe right-circularly polarised. Upon focusing through a high NA lens in the focal plane the longitudinal spin associated with the two opposite circularly polarised components sum to zero, leaving an angular momentum which is purely transverse. Due to a \textit{geometric} spin Hall effect of light (SHEL)~\cite{aiello2009transverse, neugebauer2014geometric}, this non-zero transverse AM shifts the barycentre of the beam intensity, thus allowing the fact transverse AM exists to be inferred from the focal plane intensity measurement. 

Soon afterwards the first evidence of the transverse \textit{spin angular momentum} – T-SAM – of focused beams was provided by the same group~\cite{neugebauer2015measuring}. By exciting a transverse spinning electric dipole in a probe gold nanoparticle, Neugebauer et al.~\cite{neugebauer2015measuring} mapped the transverse spin momentum density of first-order radially polarised vector and linearly polarised Gaussian beams directly. 


As discussed in Section~\ref{II}, optical properties of non-paraxial fields are dual-asymmetric: the electric and magnetic contributions do not coincide with one another as they do in paraxial optics. The transverse spin of light originates from the (non-paraxial) longitudinal field component as we have seen, and thus the electric and magnetic transverse spin densities are different. The magnetic contributions to the transverse spin density of focused beams were experimentally measured by studying the magnetic dipole resonance of a $\text{SiO}_2$-coated Si nanoparticle \cite{neugebauer2018magnetic} – see Figure~\ref{fig:8}a.

Another remarkable property of the transverse spin is that it is independent of the degree of 2D polarisation of the input beam: a focused unpolarised beam has transverse spin in the focal plane \cite{eismann2021transverse, chen2021structure} – see Figure~\ref{fig:8}c. This transverse spin of unpolarised beams has been referred to as `polarisation'-independent but it is interesting to note that $\mathbf{s}_{\perp}(\text{unpolarised})\neq\mathbf{s}_{\perp}(\text{polarised})$ unlike the optical chirality of 2D unpolarised vortex beams \cite{forbes2022optical} (see Section~\ref{VII}).  Moreover, in contrast to the general nature of non-paraxial fields, the electric and magnetic contributions to the transverse spin are identical for unpolarised beams. It has even been shown that certain singular points in the transverse spin patterns of focused linearly polarised Gaussian beams –
known a $C$ points – exhibit the topology of Möbius strips \cite{bauer2015observation, bauer2016optical} (Figure~\ref{fig:8}b). 

The T-SAM of 2D circularly polarised optical vortex beams exhibits distinctive features arising from spin-orbit interactions. Specifically, the T-SAM given by Eq.~\eqref{eq:13} for a circularly polarised input beam with coefficients $\alpha = 1/\sqrt{2}$ and $\beta = i\sigma/\sqrt{2}$ at the focal plane ($z = 0$) becomes
\begin{align}
\mathbf{s_E}^{\text{T0} \times \text{L1}}_\sigma &=\frac{I_{\text{LG}}}{c\omega k}\mathbf{\hat{\phi}}\Bigl(\frac{\ell\sigma}{r}-\text{Re}\gamma\Bigr). \label{eq:15}
\end{align}

In contrast to Eq.\eqref{eq:14}, the transverse spin for input CPL vortex beams exhibits SOI through the $\ell\sigma$-dependent term (Figure~\ref{fig:7}c). These SOI lead to distinctly different spatial distributions of the transverse spin for different values of the input OAM and SAM, including topological control through $\ell$ for vortex beams. 


The spatial distributions of transverse spin density can be highly tailorable in more complex beams. It has been shown how the transverse spin depends on the handedness of the Pancharatnam topological charge and polarisation index of both HyOP and HOP beams \cite{forbes2024spin}, and that this tunability can be used to trap and spin multiple particles \cite{li2017transverse, kumar2024probing}. Chen at al.~\cite{chen2017tightly} demonstrated how to engineer diffraction-limited focal spots with controllable transverse spin orientation through manipulating the input paraxial vector beam.

Interestingly, the transverse spin of scalar optical vortices has yet to be experimentally observed, despite the fact that current experimental techniques are largely independent of the input beam structure \cite{bauer2014nanointerferometric, eismann2024nanoscale} and should, in principle, be capable of mapping the spin distribution in these more complex fields. The transverse spin profile of linearly polarised scalar vortex beams is essentially similar to that of Gaussian beams, e.g., compare Figure~\ref{fig:7}a with \ref{fig:7}b. However, the spin–orbit interaction  uniquely exhibited by circularly polarised vortices Figure~\ref{fig:7}c presents a distinctive and worthwhile avenue for future experimental investigation. Note that, since a radial vector beam is a superposition of orthogonally circularly polarised components with $\sigma = \pm 1$ and $\ell = \mp 1$, the spatial distributions shown in Fig.~\ref{fig:7}c are reproduced in the experimental observations of Fig.~\ref{fig:8}a.
\subsubsection{Orbit-Induced-Local-Spin OILS} 

\begin{figure}[!ht]
    \includegraphics[width = \linewidth]{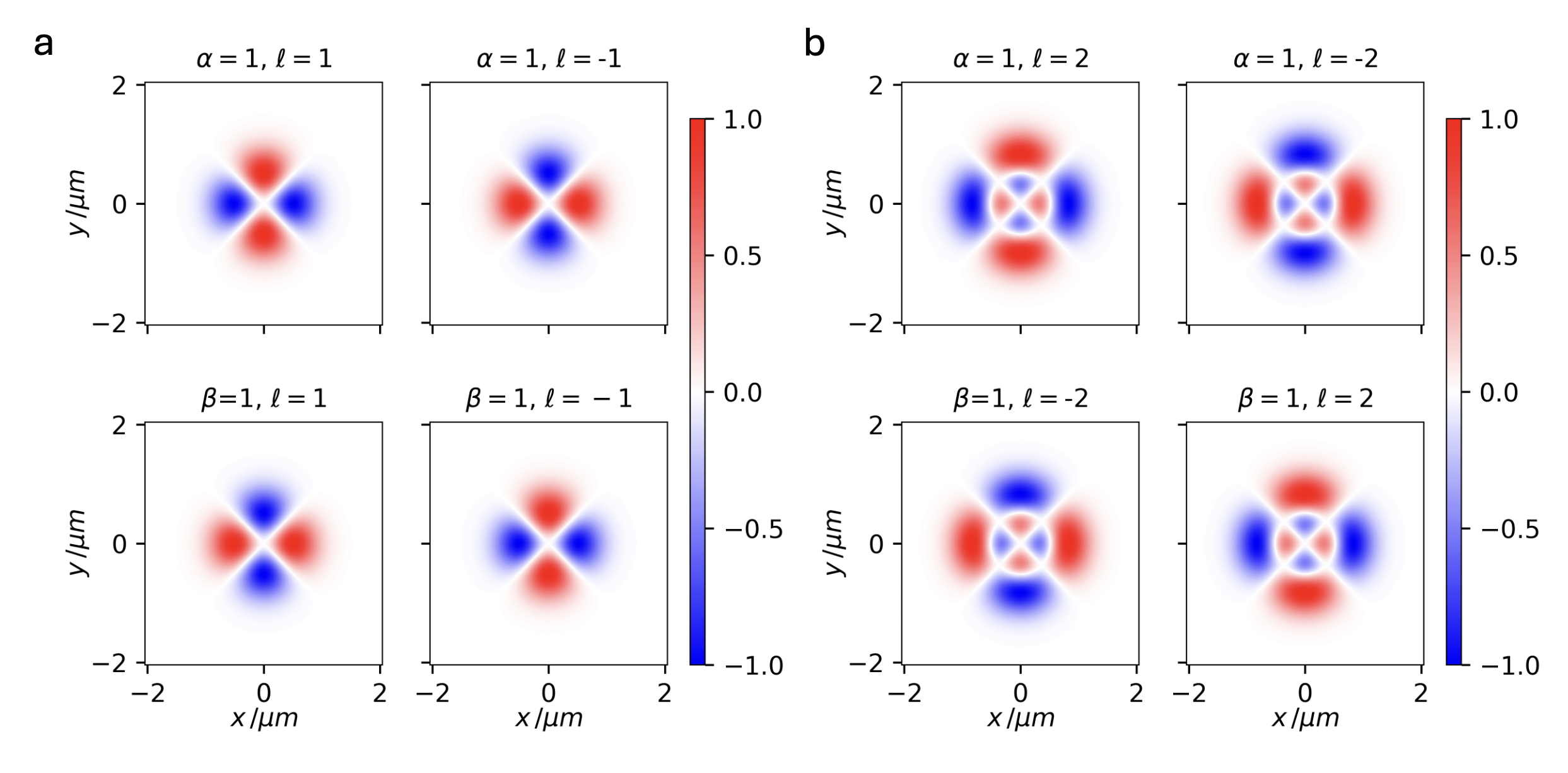}
    \caption{Spatial distributions of the L-SAM of tightly-focused 2D linearly polarised LG modes Eq.~\eqref{eq:20}: a) $|\ell|=1$; b) $|\ell|=2$. (In all plots $\lambda=w_0$ and each are individually normalised.)}
    \label{fig:9}
\end{figure}

As discussed, the $\mathbf{T_0}\times \mathbf{T_0}$ contribution to Eq.~\eqref{eq:12} generates the familiar L-SAM of paraxial beams, directly proportional to the degree of ellipticity in the $\mathbf{E}^{\text{T0}}$ field generated by the $\pi/2$ out-of-phase $x$ and $y$ components. Inspection of Eq.~\eqref{eq:8}, however, readily highlights that there in fact exists a $\pi/2$ out-of-phase relationship between the transverse components more generally. For example, for an $x$-polarised beam (i.e. $\alpha=1$), the $\mathbf{T_0}\times \mathbf{T_0}$ contribution to Eq.~\eqref{eq:12} is zero. However, there exists a $\pi/2$ out-of-phase relationship between the $x$ component of the T0 field and the $y$ component of the T2 field, thus the  $\mathbf{T_0}\times \mathbf{T_2}$ contribution to Eq.~\eqref{eq:12} generates a non-zero L-SAM, even though the paraxial field is linearly polarised. In other words, tight focusing can produce spin from spinless beams. For linearly polarised input beams (i.e. $\alpha,\beta$ are real and thus $\eta=0$), \textit{at the focal plane} $z=0$, we get for the L-SAM in terms of the Poincar\'e angle $\theta$:
\begin{align} \label{oils}
\mathbf{s_E}^{\text{T0} \times \text{T2}} &= \ell\mathbf{\hat{z}}  \frac{2I_{\text{LG}}}{c\omega k^2r}\Bigl(\text{Re}\gamma -\frac{1}{r}\Bigr)\Bigl(\sin2\theta\sin2\phi+\cos2\theta\cos2\phi\Bigr).
        \end{align}
Firstly it is clear that to be non-zero the input beams must possess OAM $\ell\neq0$. Moreover, the spatial distribution – Figure~\ref{fig:9} – rotates with the orientation $\theta$ of the input linear polarisation and the $\phi$-dependence means that, like T-SAM, there is zero total SAM generation $\int\mathbf{s}^{\text{T0} \times \text{T2}} d\mathbf{r}_\perp=0$. This requirement of OAM to generate a local, non-zero spin density, and the fact no OAM is converted to total SAM, has led to this effect being called \textit{orbit-induced-local-spin} (OILS). 


In their 2018 study, Yu et al.~\cite{yu2018orbit} used Richards-Wolf vectorial diffraction theory to model the electric field of a linearly-polarised (scalar) LG beam focused by a high-NA aplanatic lens in the focal plane. They discovered that a L-SAM density is generated for vortex beams $\ell\neq0$ that reverses sign upon changing the sign of $\ell$ and whose shape also changes upon changing the magnitude of $\ell$. Moreover, the spatial distribution of spin changes with the state of input linear polarisation, i.e. the azimuth on the Poincar\'e sphere, and that for beams with no OAM ($\ell=0$), i.e. a fundamental Gaussian beam, no SAM is generated (N.B. all of their results applied to the focal plane $z=0$). 

Kotlyar et al.~\cite{kotlyar2020spin} also studied the OILS mechanism using FDTD methods. In 2021 Forbes and Jones~\cite{forbes2021measures} discovered the $\mathbf{T_0}\times \mathbf{T_2}$ origin of the OILS mechanism using analytical theory. As well as the electric field contribution $\mathbf{s_E}^{\text{T0} \times \text{T2}}$ to OILS, this study \cite{forbes2021measures} also derived the magnetic field contribution $\mathbf{s_B}^{\text{T0} \times \text{T2}}$, highlighting that the dual OILS is zero, i.e. $\mathbf{s}^{\text{T0} \times \text{T2}}=\mathbf{s_E}^{\text{T0} \times \text{T2}}+\mathbf{s_B}^{\text{T0} \times \text{T2}}=0$ (this has important implications in forces on chiral matter, see below). An immediate consequence of the OILS effect is that it is possible to produce optical torques on absorptive dipolar particles (as long as they are not dual-symmetric) from beams which before focusing had zero spin, i.e. we get spin from spinless beams, and moreover this local spin density is controlled by the input OAM through its dependence on $\ell$. 

All of the studies above refer to OILS in scalar beams. However, there have been significantly more studies of OILS in vector beams. In 2018 three studies were published which looked at the SAM generated by tight-focusing vector vortex beams \cite{han2018catalystlike, li2018orbit, shi2018structured}, all studies utilised the Richards-Wolf framework. Han et al. \cite{han2018catalystlike} showed how radially-polarised Bessel-Gauss vector beams generate longitudinal SAM density for values of $\ell\neq0$, however in contrast to the scalar vortex beams, the L-SAM distributions are circularly-symmetric. Li et al.~\cite{li2018orbit} looked at azimuthally polarised vector vortex beams and found analogous results to those of Han et al.~\cite{han2018catalystlike}, whilst \cite{shi2018structured} studied higher-order CVVBs and discovered a similar generation of L-SAM density for input beams with OAM but with significantly more richly patterned spatial distributions. These aforementioned works, all published in 2018, were followed up with many more \cite{meng2019angular, li2021spin, geng2021orbit, fang2021photoelectronic, zhang2022ultrafast, man2022polarization, liu2024manipulation, wu2024controllable, forbes2024spin}. 

The generation of L-SAM from linearly polarised vortex beams – both scalar and vector – has traditionally been ascribed to a single optical spin–orbit interaction mechanism (i.e. OILS), arising from non-paraxial field components in tightly focused beams. However, a more detailed analysis reveals two fundamentally distinct mechanisms. We introduce the terms \emph{intrinsic OILS} (iOILS) and \emph{extrinsic OILS} (eOILS) to delineate these effects. The iOILS mechanism, rooted directly in Maxwell’s equations, arises from second-order transverse field components ($\mathbf{T}_2$) and applies to both scalar and vector vortex beams under tight focusing. It generates longitudinal spin through the interaction $\mathbf{T}_0 \times \mathbf{T}_2$. In contrast, eOILS is exclusive to vector vortex beams and does not require tight focusing. It originates from the fact that such beams are not eigenmodes of free-space propagation: their constituent spin components accumulate distinct Gouy phases, $
\zeta_\text{A(B)}[z] = \left(2p_\text{A(B)} + |\ell_\text{A(B)}| + 1\right)\arctan(z/z_\text{R}),
$  and radial amplitudes $\propto(r/w)^{|\ell_\text{A(B)}|}$ ,
leading to a spatial rotation and separation of their amplitude distributions \cite{mkhumbuza2024stokes, mkhumbuza2025topological}. This induces local variations in the paraxial spin $\mathbf{T}_0 \times \mathbf{T}_0$ across the beam profile. These distinctions are fully captured in the non-paraxial theory of cylindrical vector vortex beams~\cite{forbes2024spin}. Across both HOP and HyOP beams, the paraxial eOILS contribution dominates under most realistic conditions. This reinterpretation of OILS into iOILS and eOILS has recently gained experimental support through the observed influence of the Pancharatnam phase on the propagation dynamics of CVVBs. To date, the iOILS mechanism – and by extension, its nanoscale manifestations  remains unobserved in experiment.

Finally, L-SAM is generally associated with optical helicity (or chirality) in the paraxial regime, since the helicity density can be defined as the projection of the SAM density onto the canonical linear momentum density, i.e., $h = \mathbf{s} \cdot \mathbf{p}_{\text{o}} / p_{\text{o}}(k_z)$~\cite{bliokh2015transverse}. However, Forbes~\cite{forbes2024orbit} recently demonstrated that the L-SAM arising from iOILS – e.g., Eq.~\eqref{oils} – surprisingly carries no associated helicity (or chirality) density. It is known that the dual-symmetric SAM density contributes to a chiral radiation pressure force on dipolar chiral particles via the interference of electric-dipole magnetic-dipole transitions E1M1, given by $\mathbf{F} \propto \text{Im}[\chi]\, s_z \hat{\mathbf{z}}$~\cite{bliokh2014magnetoelectric}, where $\chi$ is the electric–magnetic dipolar polarisability. Crucially, for iOILS, the dual-symmetric L-SAM vanishes: $s_z = s_z^{E} + s_z^{B} = 0$ (as mentioned earlier \cite{forbes2021measures, forbes2024orbit}). Consequently, the chiral force also vanishes: $\mathbf{F} \propto \text{Im}[\chi]\, s_z \hat{\mathbf{z}} = 0$. This result aligns perfectly with the absence of helicity, and hence the absence of a pseudoscalar quantity required for a genuine chiral light–matter interaction. N.B. This does not rule out other forms of chiral radiation pressure that may arise from, for example, the curl of the Poynting vector~\cite{bliokh2014magnetoelectric, genet2022chiral, vernon2024decomposition}.

\subsubsection{Angular momentum from wavefront curvature} 

\begin{figure}[!ht]
    \includegraphics[width = \linewidth]{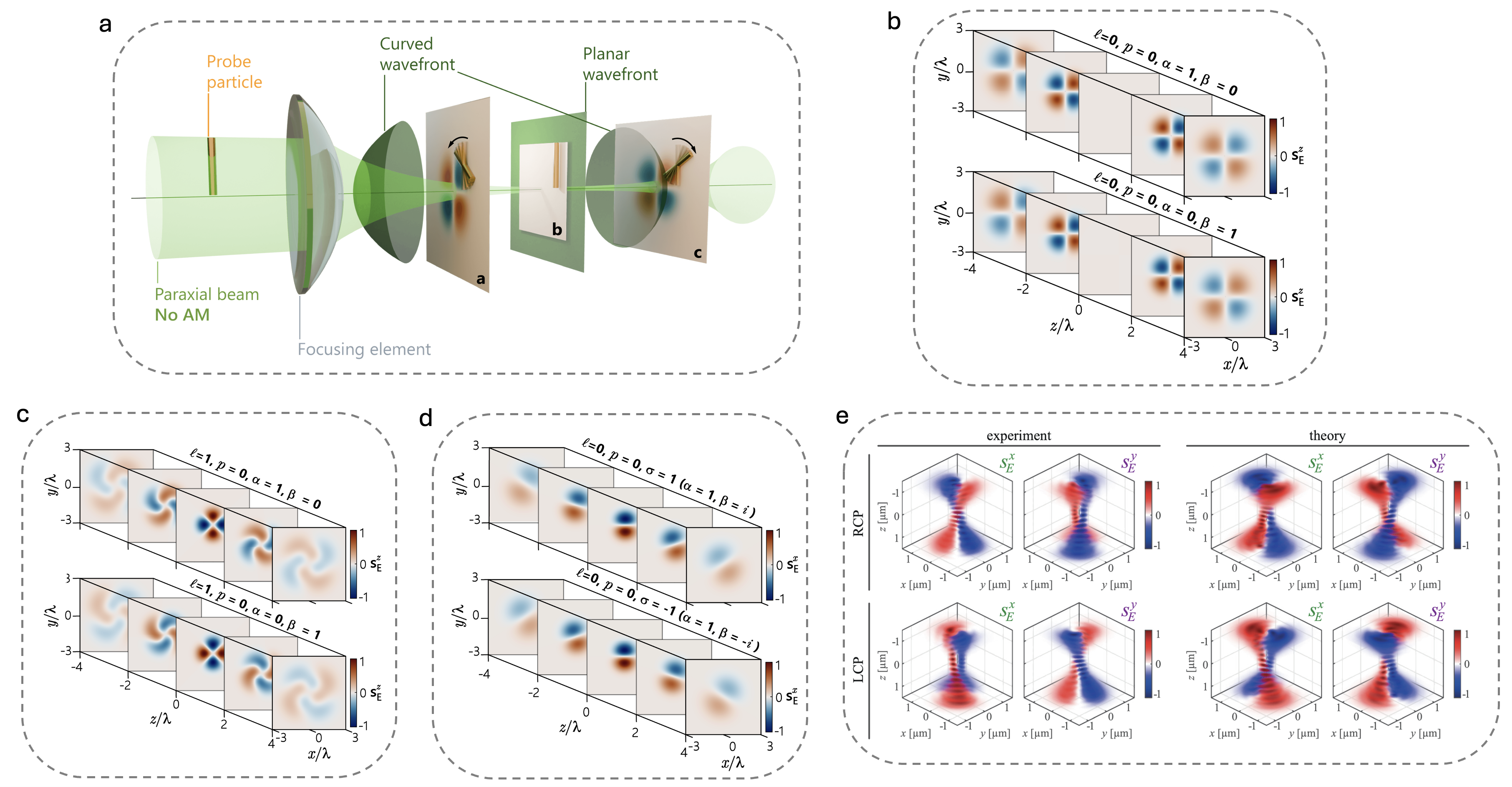}
    \caption{a) schematic highlighting how wavefront curvature can generate AM (e.g. L-SAM to spin a probe particle) from an input paraxial beam with zero AM \cite{forbes2025generating}; b) Spatial distributions of L-SAM for focused linearly-polarised Gaussian beams at different slices of $z$ \cite{forbes2025generating}; c) same as b) but for LG modes (note at $z=0$ we observe L-SAM due to iOILS) \cite{forbes2025generating}; d) Helicity-dependent transverse spin at $z\neq0$ \cite{forbes2025generating}; e) Experimental (and theory) helicity-dependent transverse spin of focused Gaussian beams \cite{eismann2019spin}.}
    \label{fig:10}
\end{figure}

It is natural to focus attention on the focal plane ($z=0$) when analysing tightly focused beams. However, Eq.~\eqref{eq:9}, which arises from the radial gradient of the paraxial field and plays a key role in non-paraxial focusing, contains a particularly interesting term: $ikr/R[z]$, originating from the gradient of the wavefront curvature phase. This term vanishes at the focal plane and in the far-field due to the divergence of $R[z]$, but becomes significant in the vicinity of the focus. Being purely imaginary, it introduces unique phase relationships distinct from other terms in the field. To illustrate its influence, consider a simple $x$-polarised Gaussian beam with zero spin and orbital angular momentum in the paraxial regime ($\ell = 0$). From Eq.~\eqref{eq:8}, we observe that focusing introduces a $\pi/2$ phase offset between the $x$-polarised T0 component and the $y$-polarised T2 component, with an azimuthal $\sin\phi\cos\phi$ dependence. This implies that a tightly focused linearly polarised Gaussian beam can acquire local L-SAM, even though it initially possessed no AM (see Figure~\ref{fig:10}a) – an effect distinct from the iOILS mechanism discussed earlier (which requires the input beam to have OAM $\ell\neq0$). Both analytical theory and Richards–Wolf diffraction simulations confirm that the $ikr/R[z]$ term is directly responsible for generating L-SAM and OAM in beams that were initially angular-momentum-free~\cite{forbes2025generating} – Figure~\ref{fig:10}b. Moreover, it plays a central role in shaping the angular momentum landscape near the focus, introducing \textit{helicity-dependent} features in the T-SAM (such as spatial distributions which rotate with $z$) and rotating the spatial structure of iOILS distributions (Figure~\ref{fig:10}c,d). In their experiment, Eismann et al.~\cite{eismann2019spin} observed rotating spatial distributions of transverse spin for tightly-focused circularly-polarised Gaussian beams, however they interpreted the effect as a relative Gouy-phase shift between the circularly polarised transverse field and the longitudinal field carrying orbital angular momentum \cite{pang2018spinning}.


\section{Chirality and Helicity} \label{VII}

Chirality is a geometrical property of objects that cannot be superimposed on their mirror images, exemplified by our hands. This asymmetry – chiral discrimination – underpins life itself~\cite{riehl2010mirror}, influencing molecular interactions such as our body's metabolism of right-handed sugars and the homochirality of amino acids. Most drugs are chiral, with regulatory trends favoring single enantiomers over mixtures of enantiomers~\cite{mcvicker2024chirality}. Beyond biology, chirality is exploited in technologies like chiral photonics and metamaterials. The ability to sense and characterise chirality is therefore essential~\cite{wang2016optical, luo2017plasmonic, yoo2019metamaterials}.
 
Optical activity refers to phenomena arising from a system’s differing response to right- and left-handed light. Natural optical activity includes circular dichroism (absorption), optical rotation, and differential scattering (e.g. differential Raman scattering is more generally known as Raman optical activity (ROA)), all with space-odd, time-even parity observables~\cite{andrews2018quantum}. Magnetic optical activity, induced by an external field, introduces space-even, time-odd parity in analogous effects. These interactions underpin chiroptical spectroscopies, essential for probing chiral structures~\cite{barron2009molecular, berova2011comprehensive, berova2012comprehensive}.

\begin{figure}[!ht]
    \includegraphics[width = \linewidth]{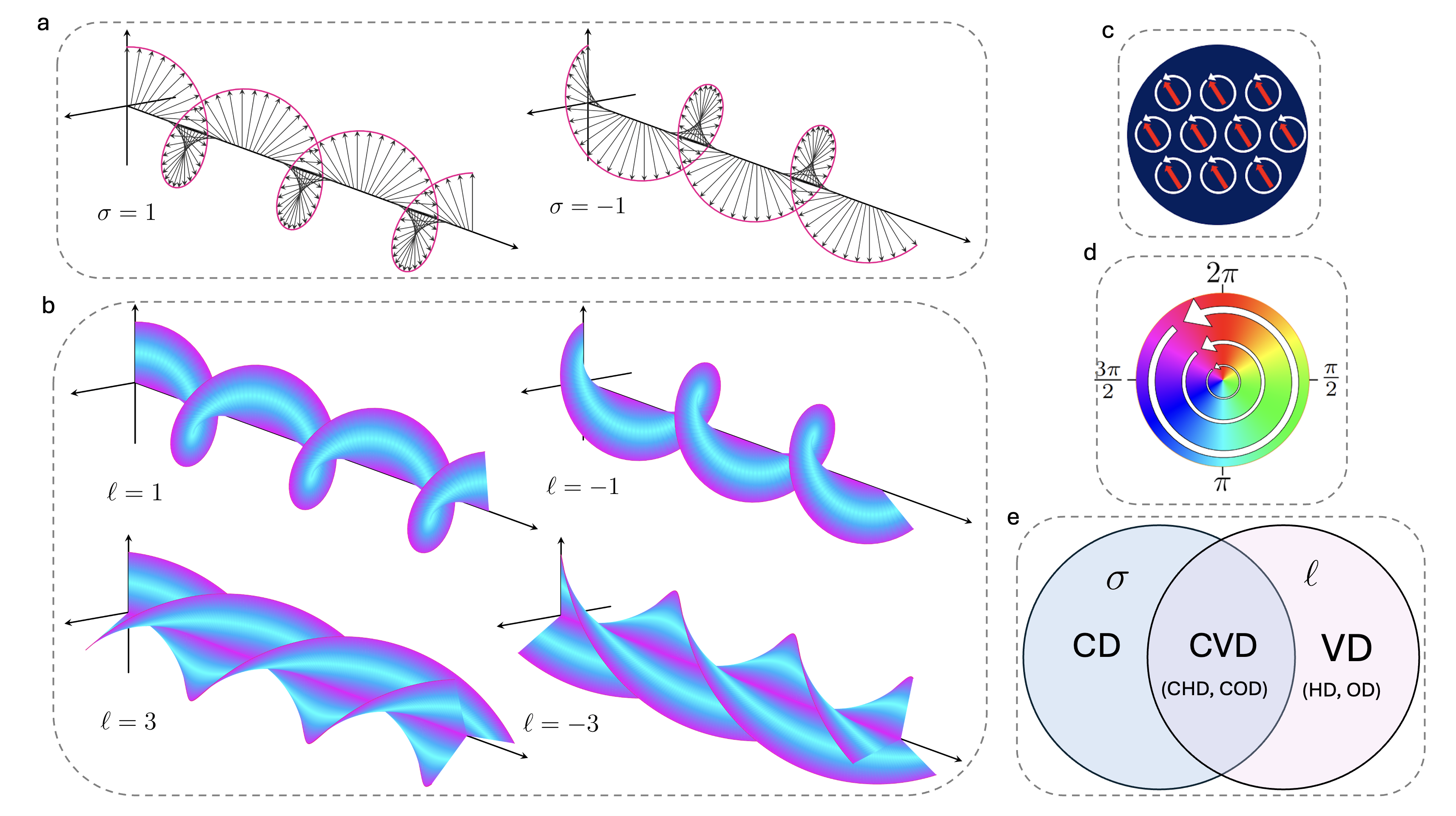}
    \caption{Chirality of structured  light: a) left $\sigma=1$ and right-handed $\sigma=-1$ circular polarisation; b) left $\ell>0$ and right-handed $\ell<0$ wavefronts of optical vortex modes; c) local nature of the chirality associated with polarisation \cite{forbes2019enhanced}; d) chirality of the wavefront (azimuthal phase structure) is a global property that spans the width of the beam \cite{forbes2019enhanced}; e) Venn diagram highlighting the various combinations chirality of polarisation and wavefront in single-photon absorption as an example. Chirality due to polarisation $\sigma$ underscores circular dichroism (CD); chirality due to wavefront $\ell$ leads to vortex dichroism (VD), or the synonymous helical/orbital dichroism (HD/OD); beams can possess both $\sigma$ and $\ell$ leading to circular-vortex dichroism (CVD).}
    \label{fig:11}
\end{figure}

Material chirality has been routinely probed optically for decades~\cite{berova2011comprehensive, berova2012comprehensive}. Chiral discrimination in light-matter interactions requires both chiral light and a chiral material. Traditionally, elliptically polarised plane waves – most commonly circularly polarised light – serve as the chiral reagent. Their helical electric and magnetic fields interact asymmetrically with chiral matter, but because the helical pitch is on the wavelength scale, molecular interactions are weak. The resulting signal, subtly altered by chirality, allows molecular handedness and absolute structure to be determined with the aid of quantum chemistry~\cite{polavarapu2016chiroptical}. The twisted wavefront of an optical vortex beam (see Figure~\ref{fig:11}b) renders them geometrically chiral: $\ell>0$ beams twist to the left, $\ell<0$ beams twist to the right. As the polarisation is independent of the wavefront, this phase chirality is distinct from the chirality associated with polarisation.
An important electromagnetic field quantity (conserved in free space) is the \textit{optical} chirality density $C$ \cite{bliokh2011characterizing, cameron2012optical}, 
\begin{align}
C=-\frac{\epsilon_0\omega}{2}\text{Im}(\mathbf{E^*\cdot\mathbf{B}}).
\label{eq:20}
\end{align}
For monochromatic beams, optical chirality is directly proportional to optical helicity $h$ ~\cite{mackinnon2019differences}, and the terms can be used interchangeably in this review. Optical chirality $C$ (Eq.~\eqref{eq:20}) and geometrical chirality are distinct; the former is only applicable to electromagnetic fields, the latter when applied to beams has been termed `Kelvin's chirality'~\cite{nechayev2021kelvin}. Geometrical chirality is binary: an object is either chiral or not. A circularly polarised and elliptical polarised plane-wave are both geometrically chiral, but a circularly polarised plane-wave has a larger \textit{optical chirality} $C$; conversely our hands are chiral yet it would be nonsensical to ascribe to them a value of optical chirality. Optical chirality quantifies light's chirality and drives electric-dipole magnetic-dipole (E1M1) contributions to linear optical activity (e.g., circular dichroism and optical rotation)~\cite{cameron2017chirality}, but does not account for all chiral light-matter interactions, such as electric-dipole electric quadrupole E1E2 contributions~\cite{craig1998molecular, barron2009molecular}. Despite not describing all chiral interactions or directly linking to geometrical chirality, optical chirality $C$ is an experimentally measurable quantity that reflects the strength of chiral light-matter interactions for small chiral `dipolar' structures.

For plane waves and paraxial beams, the optical chirality is directly proportional to the degree of ellipticity $C\propto \sigma I_\text{LG}$, taking on its maximum values for states of circular polarisation $\sigma=\pm1$, decreasing as the degree of ellipticity is reduced $-1<\sigma<1$, finally becoming zero for linearly polarised light $\sigma=0$. Although LG modes were studied by laser scientists since the 1960s~\cite{siegman2000laser}, the seminal paper on the OAM of LG modes \cite{allen1992orbital} was responsible for highlighting the chiral wavefront of such modes to a significantly broader audience. Soon researchers asked the question: can the helical wavefront of vortex modes engage in chiral light-matter interactions in an analogous fashion to the helical structure of circularly polarised light? That is to say, for example, just as circular dichroism (CD) corresponds to the differential absorption of left- $\sigma=1$ and right-handed $\sigma=-1$ circularly polarised beams, can chiral molecules differentially absorb left- $\ell=1$ and right-handed $\ell=-1$ optical vortices – vortex dichroism? (N.B. that vortex dichroism is equivalent to helical dichroism or orbital dichroism, phrases also used in the literature, we prefer the former as it is the most precise descriptor: CPL is likewise helical in structure; spin dichroism is not used because optical activity is not related to angular momentum). The motivation is threefold: exploiting $\ell$ allows larger signals (as it takes arbitrary integers unlike $\sigma$); enables access to new structural information via strong field gradients driving forbidden transitions; and supports chiroptical techniques independent of polarisation ellipticity.

Interestingly, a personal memoir by L. Allen~\cite{allen2017orbital} provides information on the first study looking at whether wavefront chirality of vortex beams engages in optical activity. Allen and M. J. Padgett shone an unpolarised OAM beam ($\sigma=0, \ell\neq0$) through sugar solutions and saw no optical rotation, thus proving the independence of polarisation and orbital angular momentum. Optical rotation by chiral molecules in a solution is purely an E1M1 effect, and thus directly proportional to $C$. It is clear that, in the paraxial experiment of Allen and Padgett, $\sigma=0$ for unpolarised paraxial light and this why they saw no optical rotation as $C=0$.

The first published study on optical activity and structured light came in 2004 by Andrews et al.~\cite{andrews2004optical}, who investigated whether the chiral wavefront of optical vortex beams could generate dichroism in chiral molecules, similar to circular polarisation in circular dichroism. They found that there is no $\ell$-dependent chiral effect with optical vortex beams. Like optical rotation, circular dichroism in chiral solutions is an E1M1 effect and proportional to $C$. This theoretical result was confirmed experimentally, showing no $\ell$-sensitivity in the circular dichroism of circularly polarised LG modes~\cite{araoka2005interactions}.

All studies discussed thus far operate within the paraxial regime, where the longitudinal field components $E_z$ and $B_z$ are neglected.  In this limit, the only dependence on the sign of the topological charge $\ell$ appears in the azimuthal phase factor $\text{exp}(i\ell\phi)$. Taking the inner product in Eq.~\eqref{eq:20} $C\propto\mathbf{E}^{*\text{T0}}\cdot\mathbf{B}^{\text{T0}}$ yields $\exp(-i\ell\phi)\cdot\text{exp}(i\ell\phi)=1$, trivially eliminating all $\ell$-dependence. Consequently, the optical chirality $C$ of paraxial vortex beams is independent of the wavefront chirality.

Chirality is inherently scale-dependent, and efficient chiral discrimination requires spatial overlap between the chiral fields and the chiral matter. Polarisation, being a local and wavelength-scale property (Figure~\ref{fig:11}c), interacts weakly with small chiral molecules or nanostructures explaining the generally weak optical activity in such systems. In contrast, the chirality encoded in the wavefront structure of vortex beams is a global property (Figure~\ref{fig:11}d), extending over the beam waist $w$, which is typically much larger than the wavelength in the paraxial regime. As a result, the mismatch in scale leads to poor overlap and weak chiral interactions.

However, tightly focusing the beam reduces the waist to near-wavelength dimensions, bringing the helical phase structure into closer correspondence with the size of chiral dipoles. As should be clear by now, such focusing introduces significant non-paraxial field components. Inspection of the higher-order terms (e.g., L1, T2) in Eq.~\eqref{eq:8} reveals explicit $\ell$-dependence beyond the azimuthal phase. This aligns with the intuition that tighter focusing enhances chiral discrimination by improving spatial overlap and introducing richer field structures that are inherently sensitive to the beam’s azimuthal phase.

\begin{figure}[!ht]
    \includegraphics[width = \linewidth]{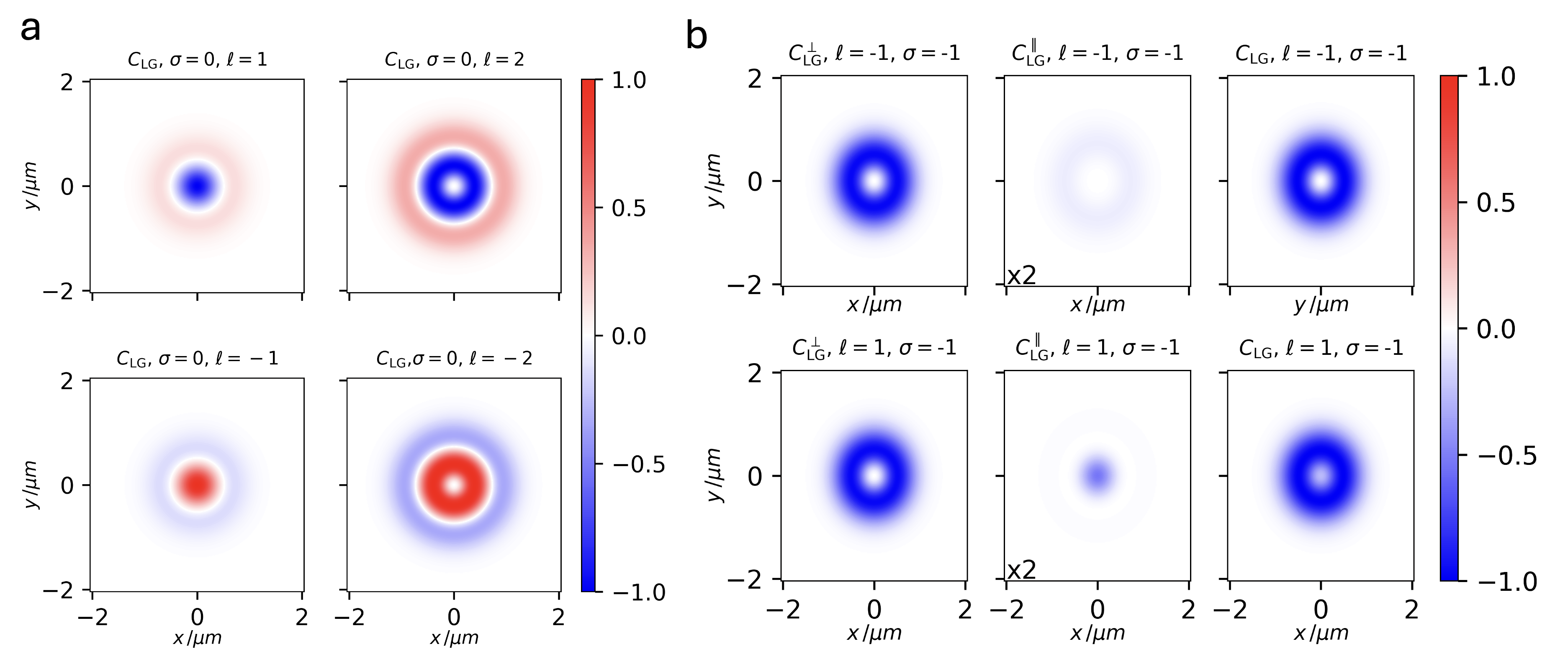}    \caption{Optical chirality density distribution of focused LG beams Eq.~\eqref{eq:21}: a) polarisation-independent $\sigma=0$ optical chirality density; b) Optical chirality of a right-handed $\sigma=-1$ circularly-polarised $\ell=\pm1$ LG beams. First column plots the transverse field contribution; middle column longitudinal field contribution; final column total $C$. N.B. the on-axis chirality density for the anti-parallel combination of $\ell$ and $\sigma$, analogous to the SOI observed in the intensity Section~{IV}. (In all plots $\lambda=w_0$ and each are individually normalised)}
    \label{fig:12}
\end{figure}

The optical chirality density of focused LG beam up to second-order in the paraxial parameter can be calculated by inserting the electric field Eq.~\eqref{eq:8} and the magnetic field (see \cite{forbes2025generating}, e.g., for the explicit form) into Eq.~\eqref{eq:20} to give (this result was first derived in \cite{forbes2021measures, green2023optical}),
\begin{align}
C &= \frac{I_{\text{LG}}\omega}{c^2}\,\Re\Biggl[\sigma +\frac{\sigma}{k^2}\Bigl(\frac{|\gamma|^2}{2}+\frac{\gamma}{r}+\gamma^{'}+\gamma^2-\frac{\ell^2}{2r^2}\Bigr)-\frac{\ell\gamma}{k^2r}\Biggr]\,.
\label{eq:21}
\end{align}
The first term in square brackets comes from the $\mathbf{E}^\text{T0}\cdot\mathbf{B}^{*\text{T0}}$ contribution, i.e. that stemming from the paraxial field, to the optical chirality density. The additional terms stem from the higher-order field components, $\mathbf{E}^\text{L1}\cdot\mathbf{B}^{*\text{L1}}$, $\mathbf{E}^\text{T0}\cdot\mathbf{B}^{*\text{T2}}$, and $\mathbf{E}^\text{T2}\cdot\mathbf{B}^{*\text{T0}}$. Remarkably, one of the pure longitudinal field components $\mathbf{E}^\text{L1}\cdot\mathbf{B}^{*\text{L1}}$ contribute a term which is linearly dependent on $\ell$, i.e. the wavefront handedness of an optical vortex beam (final term in square brackets in Eq.~\eqref{eq:21}). Moreover this term is $\sigma$-independent, and manifests for beams of light with 2D linear polarisation. In fact, as we will discuss further down, this term is independent of the 2D polarisation state of the focused beam, even persisting for unpolarised vortex beams. This $\ell$-dependent term is zero when integrated across the beam profile (Figure~\ref{fig:12}a), so does not contribute to the integral optical chirality. The spatial distributions of $C$ in the focal plane are given in Figure~\ref{fig:12}. 

The first study to highlight the remarkable role of longitudinal fields in enhancing optical chirality examined a co-propagating coherent superposition of linearly polarised Bessel beams with topological charges $\ell=1$ and $\ell=-1$ \cite{rosales2012light}. Focusing on the beam centre ($r=0$), where only longitudinal electric and magnetic fields remain, the authors demonstrated that by tuning the relative phase, polarisation, and amplitude of the two input beams, one can induce a $\pi/2$ phase shift between $E_z$ and $B_z$. This leads to localised regions of enhanced optical chirality density $C$ surpassing that of circularly polarised light, and predicts stronger chiral light–matter interactions in such engineered field configurations.

\begin{figure}[!ht]
    \includegraphics[width = \linewidth]{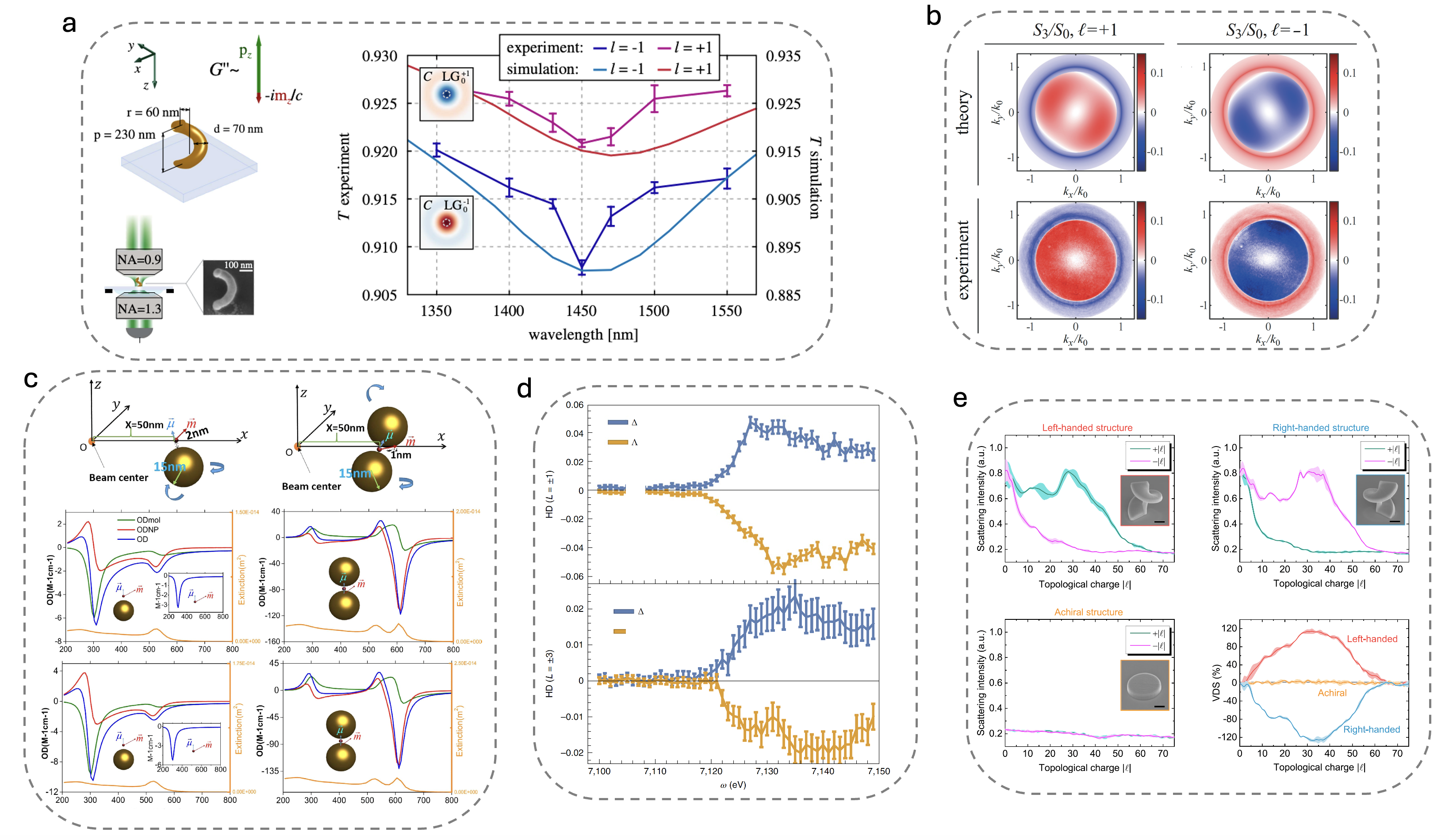}
    \caption{a) vortex dichroism in a chiral gold nanohelix \cite{wozniak2019interaction}; b) Measurement of the third Stokes parameter of the scattered light from a silicon nanoparticle in the far field allows the near-field $C$ of a focused vortex beam to be mapped \cite{nechayev2019orbital}; c) plasmon-induced vortex dichroism of chiral molecules sandwiched between gold nanoparticles \cite{wu2015plasmon}; d) vortex dichroism of chiral organometallic molecular complexes in the hard x-ray regime \cite{rouxel2022hard}; e) gigantic differential vortex scattering by chiral microstructures highlighting the scale-dependent nature of the chirality associated with vortex beams \cite{ni2021giant}.}
    \label{fig:13}
\end{figure}

A numerical study utilizing the T-matrix method by Wu et al.~\cite{wu2015plasmon} highlighted plasmon-induced vortex dichroisms of chiral molecules (Figure~\ref{fig:13}c). In particular, a chiral molecule sandwiched between two spherical gold nanoparticles exhibited significantly enhanced vortex dichroism compared to a molecule in isolation or next to a single nanoparticle. Moreover, the nanoparticles themselves also yield a vortex dichroism even though achiral. Strikingly, the signal is directly proportional to the optical chirality density Eq.~\eqref{eq:21} (when $\sigma=0)$, such that it increases with the magnitude of $\ell$ and survives orientational averaging of the particles but is zero upon spatial averaging. 

In 2019 Woz\'niak et al. \cite{wozniak2019interaction} provided the first experimental proof of the optical chirality density of a 2D linearly polarised optical vortex beam (Figure~\ref{fig:13}a). In their experiment they tightly-focus $x$-linearly polarised $\ell=\pm1, p=0$ LG modes with a resonant excitation wavelength onto a right-handed gold nanohelix which acts as a sub-wavelength-sized chiral dipolar scatterer. Measuring the transmitted light they clearly observed a difference for $\ell=1$ versus $\ell=-1$ beams, indicating a discriminatory absorption dependent on the wavefront handedness, i.e. vortex dichroism.

Light scattering by Mie particles is strongly wavelength-dependent, and at a specific wavelength $\lambda_d$, the scattering becomes dual-symmetric – satisfying the first Kerker condition – and preserves the local helicity (chirality) of the incident field. Nechayev et al.~\cite{nechayev2019orbital} showed that when tightly focused, linearly polarised LG beams are scattered by spherical silicon nanoparticles, the scattered light at $\lambda_d$ becomes purely circularly polarised. Measuring the normalised third Stokes parameter, $S_3$, in the far field – whose handedness and spatial distribution depend on the optical chirality $C$ of the input tightly focused vortex field – allows $C$, and thus properties of the input field, to be inferred (Figure~\ref{fig:13}b).

Rouxel et al.\cite{rouxel2022hard} showed VD of a disordered powder sample of enantiopure salts of the organometallic molecular complex [Fe(4,4$^{'}$-diMebpy)3]$^{2+}$ at the iron K edge (7.1 keV) (Figure~\ref{fig:13}d). From a hard x-ray beamline, spiral Fresnel zone plates were used to produce tightly focused linearly polarised vortex beams that generated asymmetry ratios of up to 5\%, significantly larger than like-for-like CD experiments. 

The clearest evidence that vortex chirality influences chiral light--matter interactions arises in the case where beams with $\sigma = 0$ are employed, thereby isolating the role of wavefront structure. However, Eq.~\eqref{eq:21} also reveals a significant interplay between the two forms of optical handedness – polarisation and wavefront topology – under non-paraxial conditions. Using a quantum theoretical framework, Forbes and Jones~\cite{forbes2021optical} demonstrated that chiral particles experience four distinct types of dichroism under tightly focused LG beams: circular dichroism (CD), vortex dichroism (VD), and two forms of circular--vortex dichroism (CVD). Analogous to the spin--orbit interaction effects on the intensity distribution $I_\text{LG}$ discussed in Section~\ref{IV}, the spatial profile of the optical chirality density $C$ depends on whether the spin $\sigma$ and orbital $\ell$ angular momenta are aligned (parallel) or opposed (anti-parallel). As a consequence, the CVD rate in chiral particles also depends strongly on this alignment. Under conditions of extreme tight focusing, the optical chirality and resulting CVD rate for the anti-parallel configuration can exceed that of conventional CD. 

Inspection of Eq.~\eqref{eq:21} reveals a remarkable result: if we average the optical chirality density $C$ over the two orthogonal circular polarisations $\sigma = \pm1$, the final term in square brackets – previously discussed in the context of linearly polarised beams – remains. In other words, this term is entirely polarisation-independent and persists even for unpolarised light. This extraordinary finding was first reported in 2022~\cite{forbes2022optical}, with several subsequent studies~\cite{green2023optical, forbes2023customized} exploring the remarkable customisation and tunability of $C$ in non-paraxial vortex beams. The spatial distribution and magnitude of $C$ were shown to depend on: (1) both the sign and magnitude of the topological charge; (2) the sign and degree of ellipticity; (3) the degree of polarisation $P$ of the input beam, including the case $P = 0$; (4) the degree of focusing; and (5) the radial index $p$. Moreover, it has even been demonstrated that unpolarised vortex beams can induce enantioselective optical forces, enabling the separation of chiral particles via a chiral gradient tweezing mechanism~\cite{forbes2022enantioselective}. 

Mayer et al.~\cite{mayer2024chiral} theoretically demonstrated that the chirality of vortex beams can be embedded into so-called synthetic chiral light~\cite{ayuso2019synthetic}, enabling enantiosensitive high-harmonic generation driven purely by electric-dipole interactions. Notably, this effect does not depend on the sign of the orbital angular momentum $\ell$. 

Cheeseman et al.~\cite{cheeseman2025nonlinear} investigated nonlinear vortex dichroism (NVD), specifically two-photon absorption, in chiral molecules under tightly focused optical vortex illumination. In addition to a dependence on the wavefront handedness, they report the remarkable finding that the spatial distribution of NVD also varies with the orientation angle $\theta$ of an input linearly polarised beam. Even more strikingly, this interaction gives rise to novel selection rules, enabling access to molecular structural information that is entirely inaccessible via conventional excitation schemes.

Radial and azimuthal first‐order vector beams (i.e. HOPs) each have zero optical chirality $C$, since they correspond purely to transverse–magnetic (TM) and transverse–electric (TE) field distributions, respectively. Herrero‐Parareda et al. \cite{herrero2024combination} showed that when these two modes are coherently superposed into a spirally polarised beam, the resulting nonzero longitudinal field components generate on-axis optical chirality – even though each constituent mode on its own carries $
C=0$. A more general analysis of non-paraxial HOPs and HyOPs \cite{forbes2024spin} revealed intricate, spatially structured distributions of optical chirality, which could be tuned via the Pancharatnam topological charge of the beams.

As discussed in Section~\ref{VI}, under non-paraxial conditions L-SAM is no longer necessarily associated with optical helicity (chirality) \cite{forbes2024orbit}. Remarkably, the polarisation-independent optical chirality of vortex beams possesses an associated T-SAM density. This result can be intuitively understood by recalling that helicity (or chirality) is defined as the projection of the spin density onto the canonical momentum density. For an unpolarised vortex beam, the dual-symmetric spin density is given by $\mathbf{s} \propto -\hat{\boldsymbol{\phi}}\, \text{Re}\gamma/k$, i.e., it is purely transverse. (Indeed, this also holds for unpolarised Gaussian beams~\cite{eismann2021transverse}; see also Section~\ref{VI}). The canonical momentum density, meanwhile, is $\mathbf{p}_{\text{o}} \propto k\hat{\mathbf{z}} + \ell\hat{\boldsymbol{\phi}}/r$. Substituting into the definition $h = \mathbf{s} \cdot \mathbf{p}_{\text{o}} / p_{\text{o}}(k_z)$ yields the final term in square brackets of Eq.~\eqref{eq:21}, as derived directly from Eq.~\eqref{eq:20}. Thus, while transverse spin is a ubiquitous feature of structured electromagnetic fields, the remarkable ability to generate $C$ from unpolarised light is unique to vortex beams, due to their azimuthal canonical momentum component. This striking result  – that T-SAM can possess helicity, while L-SAM in certain cases does not – highlights that, contrary to conventional wisdom and textbook assertions, spin, chirality, and polarisation are not intrinsically interdependent \cite{forbes2024orbit, forbes2025generalized}. Nowhere is this decoupling more evident than in non-paraxial vortex beams.

It would be remiss not to mention a distinct but complementary strand of research on optical activity involving vortex beams – one that does not directly concern the optical chirality density $C$. As noted earlier, chiral light–matter interactions can also arise from electric quadrupole $Q_{ij}$ (E2) and other higher-order multipolar couplings, which are not captured by $C$. In particular, significant attention has been directed towards electric quadrupole mechanisms in chiral interactions mediated by vortex beams. This line of inquiry began in 2016, when Brullot et al.~\cite{brullot2016resolving} demonstrated that enantiomers of phenylalanine, adsorbed onto nanoparticle aggregates, exhibited differential transmission for linearly polarised optical vortices with $\ell>0$ and $\ell<0$, while the response vanished for $\ell=0$ beams. The observed discrimination was attributed to electric quadrupole interactions between the structured light and the molecule–nanoparticle composite. Building on this, Forbes and Andrews~\cite{forbes2018optical} applied quantum electrodynamics to elucidate the fundamental role of the electric quadrupole interaction. Unlike electric dipoles, which couple directly to the electric field, quadrupoles interact with the field gradient – specifically, through terms of the form $Q_{ij} \nabla_i E_j$. In the case of vortex beams, the azimuthal gradient imparts a linear dependence on $\ell$ even within a paraxial regime, via the relation:
\begin{align}
\Bigl(\frac{\partial}{\partial r}+\frac{1}{r} \frac{\partial}{\partial \phi}+\frac{\partial}{\partial z}\Bigr) E^{\text{T0}} = \Bigl(\gamma+\frac{i\ell}{r}+ik\Bigr) E^{\text{T0}}.    
\end{align}
However, for a strong effect to manifest, spatial matching between the field structure and the molecular or nanoparticle dimensions remains crucial, and the interaction generally remains weak unless the beam is tightly focused. Remarkably, inspection of Eq.~\eqref{eq:8} shows that the longitudinal gradient of the $\mathbf{E}^{\text{L1}}$ field generates contributions of the same magnitude as the transverse gradient of the paraxial field: e.g. for $\alpha=1$,
\begin{align}
\frac{\partial}{\partial z} E^{\text{L1}}=\Bigr(\frac{i\ell}{r}\sin\phi-\gamma\cos\phi\Bigr)E^{\text{T0}}.
\end{align}
That is to say, the longitudinal gradient of the longitudinal field is of the same importance as the transverse gradient of the paraxial field in vortex beams and should not be neglected. Since these foundational studies, there has been a growing body of work dedicated to probing the chirality discrimination capabilities of vortex beams, with numerous investigations reporting enantioselective responses across a variety of systems~\cite{mullner2022discrimination, fanciulli2022observation, forbes2019enhanced, ni2021gigantic, reddy2018interaction, ji2024observation, kerber2018orbital, forbes2019raman, limstrong, sirenko2019terahertz, fanciulli2025magnetic}. For broader overviews and context, see the comprehensive reviews~\cite{forbes2021orbital, porfirev2023light}.

Regardless of the underlying electromagnetic mechanism – E1M1, E1E2, or higher – the key to harnessing the wavefront chirality of vortex beams lies in matching the spatial scale of the beam to that of the target system, e.g. the giant vortex differential scattering (120\%) of chiral microstructures \cite{ni2021gigantic} – see (Figure~\ref{fig:13}e) – and vortex dichroism (20\%) exhibited in magnetic vortices \cite{fanciulli2022observation}. This distinction arises from the scale-dependent nature of chirality: polarization-based chirality is local and fixed by the optical wavelength, whereas phase- or wavefront-based chirality spans the entire beam profile and is more readily tunable to different length scales and applications.


\section{atom optics} \label{VIII}

The transfer of angular momentum to an atom can be separated into two components: electronic motion and centre-of-mass motion. For standard electric dipole transitions, SAM is transferred to the electron, underpinning familiar selection rules, while OAM is imparted to the atom’s centre of mass~\cite{babiker2019atoms}. Babiker et al.~\cite{babiker2002orbital} first demonstrated that OAM can only couple to internal electronic degrees of freedom via higher-order multipolar transitions – the lowest order being the electric quadrupole. This is because, as discussed in Section~\ref{VII}, the quadrupole couples to the gradient of the field, with the azimuthal gradient specifically encoding the OAM.

\begin{figure}[!ht]
    \includegraphics[width = \linewidth]{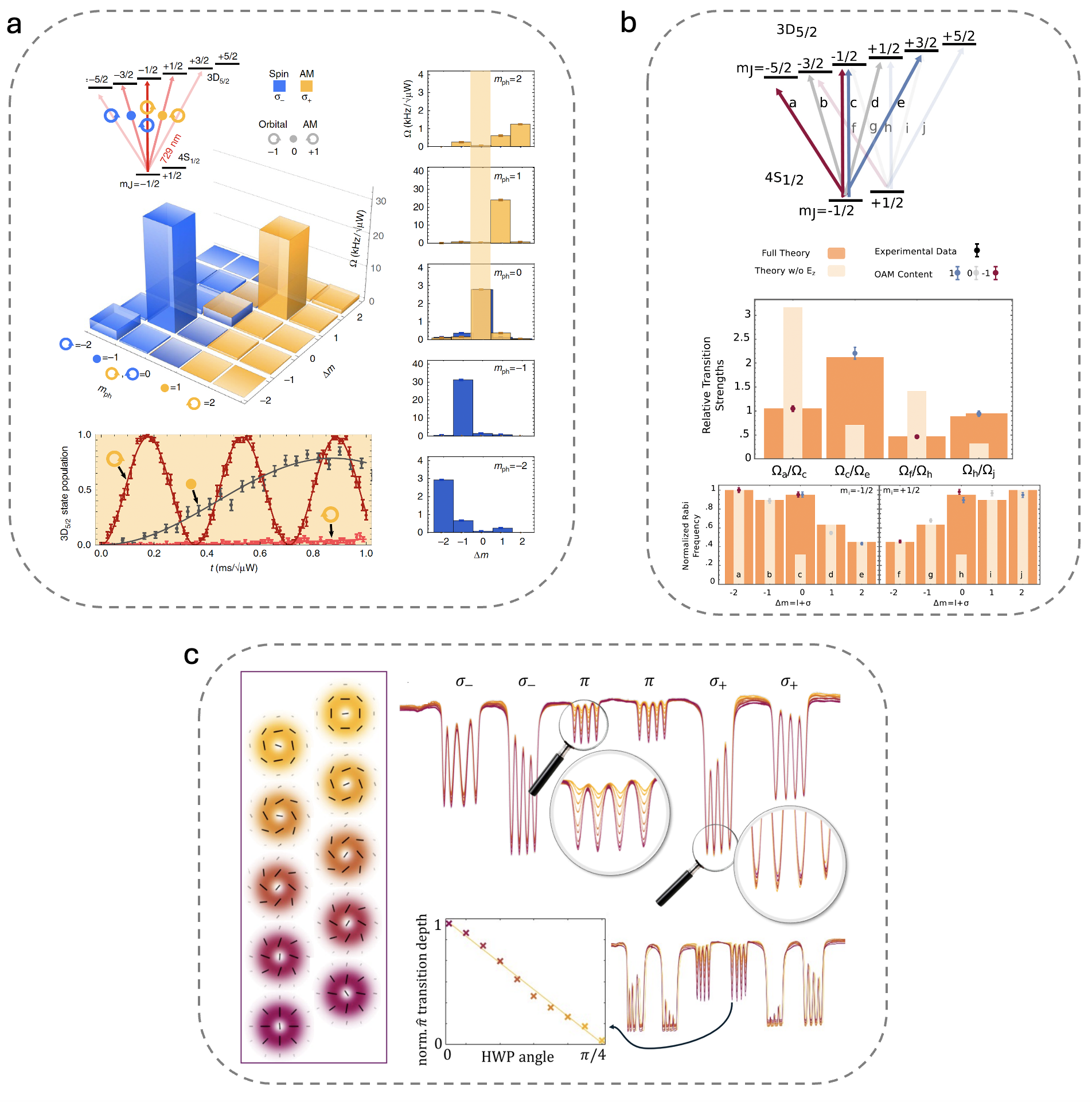}
    \caption{ a) transfer of OAM from a focused LG to laser-cooled $^{40}$Ca$^{+}$ ion \cite{schmiegelow2016transfer}; Theoretical analysis \cite{quinteiro2019reexamination} highlighting the importance of the L1 longitudinal field component in explaining the transfer of OAM in the experiment \cite{schmiegelow2016transfer};~c) driving $\pi$ transitions of the D2 line in heated \(^{87}\mathrm{Rb}\) atoms through L1 longitudinal field components \cite{svensson2025visualizing}. As the beam becomes `less radial' and 'more azimuthal`, the electric L1 component gets weaker in magnitude and consequently the $\pi$ transition strength becomes weaker.}
    \label{fig:14}
\end{figure}

The first experimental confirmation of OAM transfer to a bound electron came in 2016 by Schmiegelow et al.~\cite{schmiegelow2016transfer} (Figure~\ref{fig:14}a). A single laser-cooled $^{40}$Ca$^{+}$ ion was trapped in a segmented Paul trap, and the $4^{2}S_{1/2} \leftrightarrow 3^{2}D_{5/2}$ electric quadrupole transition was driven using focused Laguerre--Gaussian beams ($w_0 = 2700$\,nm, $\lambda \approx 729$\,nm) with $\ell = 0, \pm1$, and $\pm2$ at $r = 0$. Coherent Rabi oscillations were observed for input beams whose total AM (SAM + OAM) matched the change in the magnetic quantum number $\Delta m$, confirming OAM transfer for $\ell =  \pm1$.

As discussed in Section~\ref{IV}, focused $\ell = \pm1$ beams retain nonzero intensity on-axis, and the beam waist $\lambda/w_0 \approx 0.3$ used in the experiment by Schmiegelow et al.~\cite{schmiegelow2016transfer} was critical to observing the effect. A follow-up theoretical analysis by Quinteiro and colleagues~\cite{quinteiro2019reexamination} emphasized the importance of the L1 longitudinal field component: neglecting it led to discrepancies as large as 11 standard deviations compared to the measured values (Figure~\ref{fig:14}b). Such a result is in agreement with the discussion on quadrupoles in Section \ref{VII}. Transitions for $\ell = \pm2$ were not observed due to a vanishing on-axis intensity under the focusing conditions used – though, as noted in Section~\ref{IV}, $\ell = \pm2$ beams \emph{can} theoretically exhibit central intensity under extreme focusing. Both experimental \cite{afanasev2018experimental} and theoretical studies \cite{ramakrishna2022photoexcitation, verde2023trapped, alharbi2023significance,  mashhadi2024quadrupole} have since followed corroborating these findings. The $^{2}S_{1/2} \leftrightarrow ^{2}F_{7/2}$ electric octupole transition in $^{171}$Yb$^{+}$ has also shown to be driven by LG beams \cite{lange2022excitation}.

Radial vector beams correspond to transverse magnetic modes, while azimuthal vector beams correspond to transverse electric modes. By adjusting the input beam’s \textit{degree of radial polarisation}, one can control the strength of the longitudinal electric field, which diminishes as the beam shifts toward azimuthal polarisation. Leveraging this, Svensson et al.~\cite{svensson2025visualizing} directly investigated the L1 electric field component of tightly focused vector beams via \(\pi\) transitions of the D2 line in heated \(^{87}\mathrm{Rb}\), using a single-beam geometry. Specifically, it was observed that the magnitude of the \(\pi\) transition decreases as the beam’s “radiality” (i.e., the L1 field component) is reduced – see Figure~\ref{fig:14}c.

\section{Thoughts on the future} \label{IX}

The science of tightly focused optical vortex beams is entering a new era, marked by fundamental discoveries and rapidly expanding applications. These beams, which carry OAM, undergo profound transformations when brought into tight focus, giving rise to rich and unexpected optical phenomena. Unlike in the paraxial regime – where spin and orbital AM are separable and conserved – tight focusing leads to strong spin–orbit coupling, nontrivial longitudinal field components, and novel forms of structured angular momentum such as OILS and transversely spinning fields (Section~\ref{VI}).

One of the most exciting prospects lies in molecular and material imaging. Nonlinear microscopy, which relies on multiphoton absorption and harmonic generation, can benefit greatly from the structured focal fields of vortex beams. These beams can generate subwavelength hotspots, structured intensity distributions, and chiral-selective interactions, all of which may improve sensitivity and resolution beyond conventional limits (Sections~\ref{IV} and \ref{VII}). Tightly focused vortex beams offer unique, polarisation-independent routes to manipulate and probe molecular chirality, including in isotropic media where conventional techniques fall short. This could lead to breakthroughs in miniaturised and on-chip photonic devices for chiral sensing and diagnostics.

Theoretical work continues to uncover new physical mechanisms enabled by tight focusing. For instance, angular momentum can be generated in the focal volume even when the input beam carries none – a striking effect driven by the interplay of wavefront curvature and vectorial field structure (Section~\ref{VI}). Moreover, the optical chirality density and other conserved quantities become spatially complex in the focal region, opening new opportunities for light–matter coupling that is sensitive to both spatial and angular momentum characteristics of light (Section~\ref{VI} and \ref{VII}).

Looking to the future, nascent areas not covered in this review look set to further increase the importance of structured light at the nanoscale. Optical skyrmions present a novel approach for encoding and manipulating structured light fields through topologically protected features, offering enhanced robustness for information processing and potential for nanoscale photonic applications \cite{shen2024optical}. Recent studies have reported the emergence of a new class of optical skyrmions in tightly focused vortex beams \cite{gutierrez2021optical, mata2025skyrmionic}. 

The interaction of focused beams with complex materials, such as condensed matter systems~\cite{quinteiro2022interplay} and Mie particles, is also emerging as a key direction in the field. These systems can couple to light via more exotic electromagnetic quantities \cite{bliokh2014magnetoelectric, shi2023advances}, such as reactive helicity, the real and imaginary parts of the Poynting vector~\cite{zhou2022observation}, and the intricate higher-order field gradients inherent to focused structured beams. In parallel, substantial attention is now being devoted to the interaction between high-energy vortex photons (e.g., extreme ultraviolet and X-rays) generated by free-electron lasers and synchrotron sources~\cite{rebernik2017extreme, fanciulli2022observation, rouxel2022hard, mccarter2024generation, fanciulli2025magnetic}. These highly non-linear light–matter interactions offer considerable potential for both theoretical and experimental advances.
   
As fabrication and beam-shaping techniques advance \cite{wang2018recent, he2022towards, schulz2024roadmap} – through metasurfaces, spatial light modulators, and novel optical elements – the ability to precisely control vortex beams at the nanoscale will only improve. This paves the way for applications in optical trapping, quantum optics, and nanophotonic device integration.

In summary, tightly focused optical vortex beams are rapidly evolving from their theoretical foundations into a powerful and versatile experimental tool. They offer a new way of structuring light at the nanoscale and promise breakthroughs across disciplines – from fundamental optics to chemistry, biology, and quantum technologies. The coming years are likely to see an acceleration in both conceptual understanding and real-world exploitation, making this one of the most vibrant frontiers in modern photonics.



\section*{List of Acronyms}

\begin{ruledtabular}
\begin{tabular}{ll  ll}
AM & Angular momentum & NA & Numerical aperture \\
CD & Circular dichroism & OAM & Orbital angular momentum \\
CP & Circular polarisation & OILS & Orbit-induced local spin \\
CVD & Circular vortex dichroism & i-OILS & Intrinsic orbit-induced local spin \\
CVVB & Cylindrical vector vortex beam & e-OILS & Extrinsic orbit-induced local spin \\
E1E2 & Electric-dipole electric-quadrupole & SAM & Spin angular momentum \\
E1M1 & Electric-dipole magnetic-dipole & SOI & Spin-orbit interaction \\
ENZ & Epsilon-near-zero & STED & Stimulated emission depletion (microscopy) \\
HOP & Higher-order Poincaré & T0 & Zeroth-order transverse field \\
HyOP & Hybrid-order Poincaré & T2 & Second-order transverse field \\
L1 & First-order longitudinal field & TE & Transverse electric \\
LG & Laguerre-Gaussian & TM & Transverse magnetic \\
L-SAM & Longitudinal spin angular momentum & T-SAM & Transverse spin angular momentum \\
VB & Vector beam & VD & Vortex dichroism \\
\end{tabular}
\end{ruledtabular}

\vspace{5pt}

\bibliography{references}

\newpage
\twocolumngrid
\end{document}